\documentclass[review]{elsarticle}
\usepackage{lineno,hyperref}
\modulolinenumbers[5]

\journal{}

\makeatletter
\def\ps@pprintTitle{%
  \let\@oddhead\@empty
  \let\@evenhead\@empty
  \let\@oddfoot\@empty
  \let\@evenfoot\@oddfoot
}
\makeatother

\usepackage[ruled]{algorithm2e}
\renewcommand{\algorithmcfname}{ALGORITHM}
\SetAlFnt{\small}
\SetAlCapFnt{\small}
\SetAlCapNameFnt{\small}
\SetAlCapHSkip{0pt}
\IncMargin{-\parindent}

 \usepackage{graphicx}
 \usepackage{color,soul}
 \usepackage{framed}
 \usepackage{url}
 \usepackage{pslatex}
 \usepackage{amsmath,amssymb}
 \usepackage{subfig}
 \usepackage{setspace}
 \usepackage{multicol}
 \usepackage{multirow}
 \usepackage{caption}
 \usepackage{wrapfig}
 \usepackage{booktabs}
 \usepackage{tablefootnote}
\usepackage[framemethod=tikz]{mdframed}
\usepackage{lipsum}
\usepackage{footnote}

\newcommand{\green}[1]{{#1}}

\usepackage[font=footnotesize,labelfont=bf]{caption}

\usepackage{tikz}
\newcommand*\mycirc[1]
{\tikz[baseline=(char.base)]{
  \node[shape=circle,draw,inner sep=0.5pt] (char) {#1};}}

\newcommand{\ignore}[1]{}

\newcommand{\breakTable}[2][c]{%
  \begin{tabular}[#1]{@{}c@{}}#2\end{tabular}}

\newcommand*{\TitleFont}{%
      \usefont{\encodingdefault}{\rmdefault}{}{n}%
      \fontsize{15}{20}%
      \selectfont}

\usepackage[normalem]{ulem}

\expandafter\def\expandafter\UrlBreaks\expandafter{\UrlBreaks
  \do\a\do\b\do\c\do\d\do\e\do\f\do\g\do\h\do\i\do\j%
  \do\k\do\l\do\m\do\n\do\o\do\p\do\q\do\r\do\s\do\t%
  \do\u\do\v\do\w\do\x\do\y\do\z\do\A\do\B\do\C\do\D%
  \do\E\do\F\do\G\do\H\do\I\do\J\do\K\do\L\do\M\do\N%
  \do\O\do\P\do\Q\do\R\do\S\do\T\do\U\do\V\do\W\do\X%
  \do\Y\do\Z}

\date{}

\begin{document}

\begin{frontmatter}


\title{\TitleFont Achieving both High Energy Efficiency\\and High Performance in On-Chip Communication\\using Hierarchical Rings with Deflection Routing \vspace{0.15in}}

\author{Rachata Ausavarungnirun \quad Chris Fallin \quad Xiangyao Yu$\dagger$ \quad Kevin Kai-Wei Chang\\ [2pt]
Greg Nazario \quad  Reetuparna Das$\S$ \quad Gabriel H. Loh$\ddagger$ \quad  Onur Mutlu \\ [7pt]
Carnegie Mellon University \quad $\S$University of Michigan \quad $\dagger$MIT \quad $\ddagger$AMD\\
\vspace{-0.5in}
}

%
%
%
\vspace{-1.5in}
\begin{abstract}

\singlespacing

  Hierarchical ring networks,
  which hierarchically connect multiple levels of rings, have been
  proposed in the past to improve the scalability of ring
  interconnects, but past hierarchical ring designs sacrifice some of
  the key benefits of rings by reintroducing more complex in-ring
  buffering and buffered flow control. Our goal in this paper is to
  design a new hierarchical ring interconnect that can maintain most
  of the simplicity of traditional ring designs (i.e., no in-ring
  buffering or buffered flow control) while achieving high scalability
  as more complex buffered hierarchical ring designs.

  To this end, we revisit the concept of a hierarchical-ring
  network-on-chip.  Our design, called \textbf{HiRD} (Hierarchical
  Rings with Deflection), includes critical features that enable us to
  mostly maintain the simplicity of traditional simple ring topologies
  while providing higher energy efficiency and scalability.
  \emph{First}, HiRD does not have {\em any} buffering or buffered
  flow control within individual rings, and requires only a small
  amount of buffering between the ring hierarchy levels.  When
  inter-ring buffers are full, our design simply \emph{deflects} flits
  so that they circle the ring and try again, which eliminates the
  need for in-ring buffering.  \emph{Second}, we introduce two simple
  mechanisms that together provide an end-to-end delivery guarantee
  within the entire network (despite any deflections that occur)
  without impacting the critical path or latency of the vast majority
  of network traffic.


  Our experimental evaluations on a wide variety of multiprogrammed
  and multithreaded workloads and synthetic traffic patterns show that
  HiRD attains equal or better performance at better energy efficiency
  than multiple versions of both a previous hierarchical ring design
  and a traditional single ring design. We also extensively analyze
  our design's characteristics and injection and delivery
  guarantees. We conclude that HiRD can be a compelling design point
  that allows higher energy efficiency and scalability while retaining
  the simplicity and appeal of conventional ring-based designs.



\end{abstract}



\end{frontmatter}

\singlespacing

\linespread{0.85}
\section{Introduction}

Interconnect scalability, performance, and energy efficiency are
first-order concerns in the design of future CMPs (chip
multiprocessors). As CMPs are built with greater numbers of cores,
centralized interconnects (such as crossbars or shared buses) are no
longer scalable. The Network-on-Chip (NoC) is the most
commonly-proposed solution~\cite{packetsnotwires}: cores exchange
packets over a network consisting of network switches and links
arranged in some topology.

Mainstream commercial CMPs today most commonly use \emph{ring}-based
interconnects.  Rings are a well-known network
topology~\cite{principles,benini-nocs,hemani-nocs}, and the idea behind a ring topology is
very simple: all routers (also called ``ring stops'') are connected by
a loop that carries network traffic. At each router, new traffic can
be injected into the ring, and traffic in the ring can be removed from
the ring when it reaches its destination. When traffic is traveling on
the ring, it continues uninterrupted until it reaches its
destination. A ring router thus \emph{needs no in-ring buffering or
  flow control} because it prioritizes on-ring traffic. In addition,
the router's datapath is very simple compared to a mesh router,
because the router has fewer inputs and requires no large,
power-inefficient crossbars; typically it consists only of several
MUXes to allow traffic to enter and leave, and one pipeline register.
Its latency is typically only one cycle, because no routing decisions
or output port allocations are necessary (other than removing traffic
from the ring when it arrives). Because of these advantages, several
prototype and commercial multicore processors have utilized ring
interconnects: the Intel Larrabee~\cite{larrabee}, IBM
Cell~\cite{cell}, and more recently, the Intel
Sandy~Bridge~\cite{sandybridge}.

Unfortunately, rings suffer from a
fundamental scaling problem because a ring's bisection bandwidth does
not scale with the number of nodes in the network. Building more
rings, or a wider ring, serves as a stopgap measure but increases the
cost of every router on the ring in proportion to the bandwidth
increase. As commercial CMPs continue to increase core counts, a new
network design will be needed that balances the simplicity and low
overhead of rings with the scalability of more complex topologies.

\begin{figure}[h]
\centering
\includegraphics[width=2in]{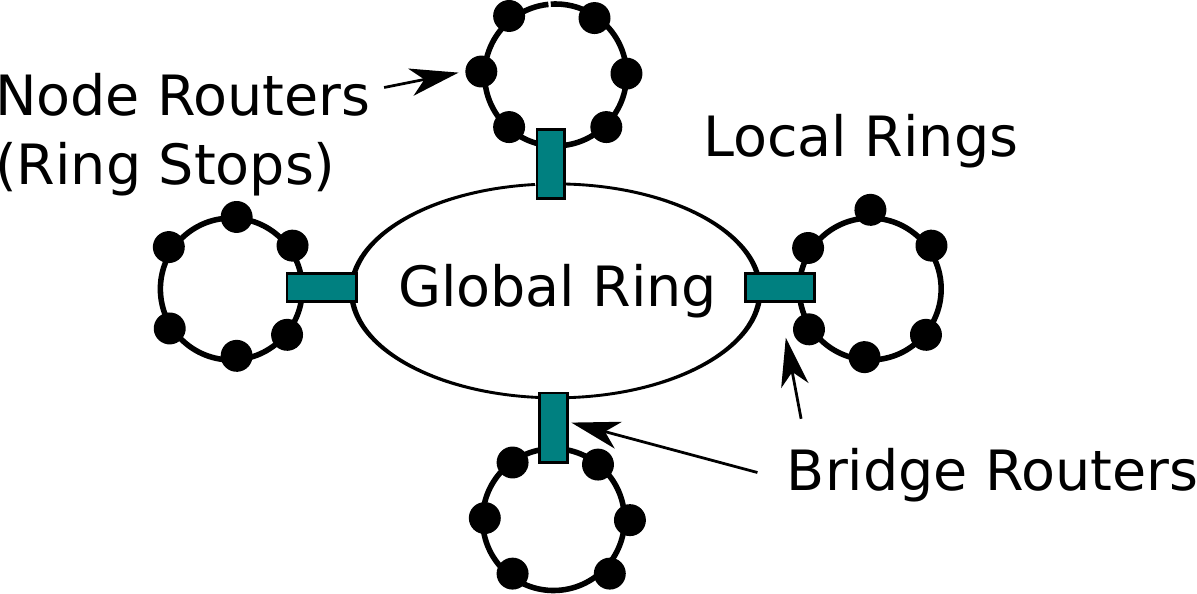}
\caption{\footnotesize \bf A traditional hierarchical ring
  design~\protect\cite{ravindran97,xiangdong95,hr-model,ravindran98,numachine}
  allows ``local rings'' with simple node routers to scale by
  connecting to a ``global ring'' via bridge routers.}
  \label{fig:hring}
\end{figure}

A hybrid design is possible: rings can be constructed in a
\emph{hierarchy} such that groups of nodes share a simple ring
interconnect, and these ``local'' rings are joined by one or more
``global'' rings. Figure~\ref{fig:hring} shows an example of such a
\emph{hierarchical ring} design. Past
works~\cite{ravindran97,xiangdong95,hr-model,ravindran98,numachine}
proposed hierarchical rings as a
scalable 
network. These proposals join rings with \emph{bridge routers}, which
reside on multiple rings and transfer traffic between rings. This
design was shown to yield good performance and
scalability~\cite{ravindran97}. The state-of-the-art
design~\cite{ravindran97} requires \emph{flow control and buffering}
at every node router (ring stop), because a ring transfer can make one
ring back up and stall when another ring is congested.  While this
previously proposed hierarchical ring is much more scalable than a single ring~\cite{ravindran97}, 
the reintroduction of in-ring buffering and flow control
nullifies one of the primary advantages of using ring networks in the
first place (i.e., the lack of buffering and buffered flow control
within each ring).

\textbf{Our goal} in this work is to design a ring-based topology that is simpler and more efficient than prior ring-based topologies. To this end, our design
uses simple ring networks that do not introduce any in-ring buffering or flow control.
Like past proposals, we utilize a hierarchy-of-rings topology to achieve higher
scalability.  However, beyond the topological
similarities, our design is very different in how traffic is handled within
individual rings and between different levels of rings.  We introduce a
new \emph{bridge router} microarchitecture that facilitates the transfer
of packets from one ring to another.  It is in these, and {\em only} these,
limited number of bridge routers where we require any buffering.


\textbf{Our key idea} is to allow a bridge router with a full buffer to \emph{deflect} packets.
Rather than requiring buffering and flow control in the ring, packets
simply cycle through the network and try again.  While
deflection-based, bufferless networks have been proposed and evaluated
in the
past~\cite{hotpotato,cm,hep,tera,casebufferless,bless_switching,chipper,hotnets2010,sigcomm12,hird,maze-routing,minbd-book}, our
approach is effectively an elegant hybridization of bufferless (rings)
and buffered (bridge routers) styles.
To prevent packets from potentially deflecting around a ring
arbitrarily many times (i.e., to prevent livelock), we introduce two
new mechanisms, the \emph{injection guarantee} and the \emph{transfer
  guarantee}, that ensure packet delivery even for
adversarial/pathological conditions (as we discuss in
~\cite{hird} and evaluate with worst-case traffic in
\S\ref{sec:eval-guarantees}).

This simple hierarchical ring design, which we call \emph{HiRD} (for
Hierarchical Rings with Deflection), provides a more scalable
network architecture while retaining the key simplicities of
ring networks (no buffering or flow control within each ring).
We show in our evaluations that HiRD provides better performance, lower power,
and better energy efficiency with respect to the \ignore{state-of-the-art }buffered hierarchical ring
design~\cite{ravindran97}.



In summary, \textbf{our major contributions} are:

\begin{itemize}

\item We propose a new, low-cost, hierarchical ring NoC design based
  on very simple router microarchitectures that achieve single-cycle
  latencies. This design, \emph{HiRD}, places an ordinary ring router
  (without flow control or buffering) at every network node, connects
  local rings with global rings using \emph{bridge routers}, which have 
  minimal buffering and use deflection rather than buffered flow control 
  for inter-ring transfers.

\item We provide new mechanisms for \emph{guaranteed delivery of
  traffic} ensuring that inter-ring transfers do not cause livelock or
  deadlock, even in the worst case.

\item We qualitatively and quantitatively compare HiRD to several
  state-of-the-art NoC designs. We show competitive performance to
  these baselines, with better energy efficiency than all prior
  designs, including, most importantly, the\ignore{state-of-the-art}
  hierarchical ring design with in-ring buffering and buffered flow
  control~\cite{ravindran97}. We conclude that HiRD represents a
  compelling design point for future many-core interconnects by
  achieving higher performance while maintaining most of the
  simplicity of traditional ring-based designs.

\end{itemize}


\ignore{

To bridge the gap, we begin with a ring-based approach, and observe
that to reduce ring contention caused by higher node counts, we can
split the network into several smaller \emph{local rings} composed of
simple ring routers. Then, using slightly more complex \emph{bridge
  routers} that join two rings, but using a smaller number of them, we
join the local rings with one or more \emph{global rings} which
transfer traffic between local rings. By adjusting the number of
global rings and their width, we can retain the bisection bandwidth of
the more complex mesh network without adopting its complex
routers. The operation of the network is only modestly more complex
than a single ring. When traffic is sent between two nodes in the same
local ring, the routers in that ring communicate as before, injecting
traffic into the local ring whenever there is a free slot and removing
traffic at its destination. However, when traffic is sent between
nodes in different local rings, the bridge router in the sender's
local ring transfers the traffic to the global ring, and the bridge
router in the receiver's local ring transfers the traffic from the
global ring into the destination ring.

While hierarchical-ring topologies have been proposed
before~\cite{ravindran97,xiangdong95,hr-model,ravindran98,numachine},
we adopt principles from deflection (hot-potato)
routing~\cite{hotpotato} in order to make the routers as simple and
energy-efficient as possible. We use minimal buffering at bridge
routers between rings, no buffering at node routers that connect
CPU/caches to the rings, and no token flow control. Instead, traffic
always moves around rings, and when a flit needs to transfer rings but
the transfer FIFO is full, the flit circles its ring and tries
again. To guarantee forward progress, we design explicit mechanisms
that detect and recover from starvation at two key points (ring
injection and ejection). We call the resulting design
the \emph{HiRD} interconnect, for ``Hierarchical Rings with
Deflection,'' and our evaluations show that \emph{HiRD} provides
better energy-efficiency and lower area cost than previous
hierarchical-ring topologies.

HiRD has multiple advantages. First, the router at each node
(i.e., processor, cache, or memory controller) is a very simple ring
router with no buffers, crossbars, or other costly components,
significantly reducing complexity, die area and power. Second, the
hierarchical structure is a good match for traffic patterns that have
some locality. Even when there is no locality (which is unlikely as
core counts continue to increase, because locality-aware mechanisms
that reduce on-chip communication overhead are becoming more critical
to maintain effective
scalability~\cite{CloudCache,das09,udipi10,cho06}), it is more
effective to add more global rings, or widen existing global rings,
than it is to widen the ring on which every local router sits, as we
show in \S\ref{sec:evaluation} by comparing to a single-wide-ring
design. Third, the global ring can act as an express network for
long-distance traffic by directly connecting two local rings without
impacting the nodes on rings between the endpoints. In addition, this
``express ring'' effect can yield lower link-traversal cost than a
single ring connecting all nodes if it leads to a more direct
traversal path (i.e., no need to visit many nodes around the
ring). Hence, relative to a traditional ring design, a hierarchical
ring architecture attains better scalability and lower cost for a
given performance target. It also uses routers that are significantly
simpler and more energy-efficient than conventional mesh routers, as
supported by our extensive experimental evaluations. \textbf{Our
contributions} are:

\begin{itemize}

\item We propose new router designs which enable a simple, low-cost,
hierarchical ring NoC.  Our proposal, \emph{HiRD}, places an ordinary ring
router at every network node, connects local rings with global rings, and joins
local and global rings using \emph{bridge routers} that have minimal
buffering, and use deflection rather than buffered flow control for inter-ring
transfers.

\item We provide new mechanisms for \emph{guaranteed delivery of
  traffic} ensuring that inter-ring transfers do not cause livelock or
  deadlock, even in the worst case.

\item We qualitatively and quantitatively compare HiRD to four\ignore{state-of-the-art}
NoC designs: (i) a buffered-router 2D-mesh interconnect, (ii) a bufferless
deflection network (CHIPPER~\cite{chipper}), (iii) a single-ring interconnect,
and (iv) a hierarchical ring that has buffers at every node and requires flow
control~\cite{ravindran97}. Although the main goal of this work is not to
comprehensively evaluate the topology design space, but rather to make a
hierarchical design simpler and more efficient, we nevertheless show
competitive performance to these baselines, with equal or better energy
efficiency than prior designs, including the hierarchical ring with in-ring
buffering.

\end{itemize}

}

\section{HiRD: Simple Hierarchical Rings with Deflection}
\label{sec:mech}

In this section, we describe the operation of our network design
\emph{HiRD}, or Hierarchical Rings with Deflection. HiRD is built on
several basic operation principles:

\begin{enumerate}

\item Every node (e.g., CPU, cache slice, or memory controller)
resides on one \emph{local ring}, and connects to one \emph{node
router} on that ring.

\item Node routers operate exactly like routers (ring stops) in a
single-ring interconnect: locally-destined flits are removed from
the ring, other flits are passed through, and new flits can inject
whenever there is a free slot (no flit present in a given
cycle). There is no buffering or flow control within any local ring;
flits are buffered only in ring pipeline registers. Node routers
have a single-cycle latency.

\item Local rings are connected to one or more levels of \emph{global
rings} to form a tree hierarchy.

\item Rings are joined via \emph{bridge routers}. A bridge router has
a node-router-like interface on each of the two rings it
connects, and has a set of transfer FIFOs (one in each direction)
between the rings.

\item Bridge routers consume flits that require a transfer whenever
the respective transfer FIFO has available space. The head flit in a
transfer FIFO can inject into its new ring whenever there is a free
slot (exactly as with new flit injections). When a flit requires a
transfer but the respective transfer FIFO is full, the flit remains
in its current ring. It will circle the ring and try again next time
it encounters the correct bridge router (this is a
\emph{deflection}).

\end{enumerate}

By using \emph{deflections} rather than buffering and blocking flow
control to manage ring transfers, HiRD retains node router simplicity,
unlike past hierarchical ring network designs. This change comes at
the cost of potential livelock (if flits are forced to deflect
forever). We introduce two mechanisms to provide a deterministic
guarantee of livelock-free operation in~\cite{hird}.

While deflection-based bufferless routing has been previously proposed and
evaluated for a variety of off-chip and on-chip interconnection networks
(e.g.,~\cite{hotpotato,casebufferless,bless_switching,chipper,minbd,hotnets2010,sigcomm12}), deflections
are trivially implementable in a ring: if deflection occurs, the
flit\footnote{All operations in the network happen in a flit level similar
to previous works~\cite{casebufferless,bless_switching,chipper,minbd,hotnets2010,sigcomm12}.} continues 
circulating in the ring.  Contrast this to past deflection-based
schemes that operated on mesh networks where multiple incoming flits may need
to be deflected among a multitude of possible out-bound ports, leading to much
more circuit complexity in the router microarchitecture, as shown
by~\cite{chipper,scarab,michelog10}.  Our application of deflection to rings
leads to a simple and elegant embodiment of bufferless routing.

\subsection{Node Router Operation}
\label{sec:mech-node-router}

\begin{figure}[ht]
\centering
\vspace{-0.1in}
\includegraphics[width=2.2in]{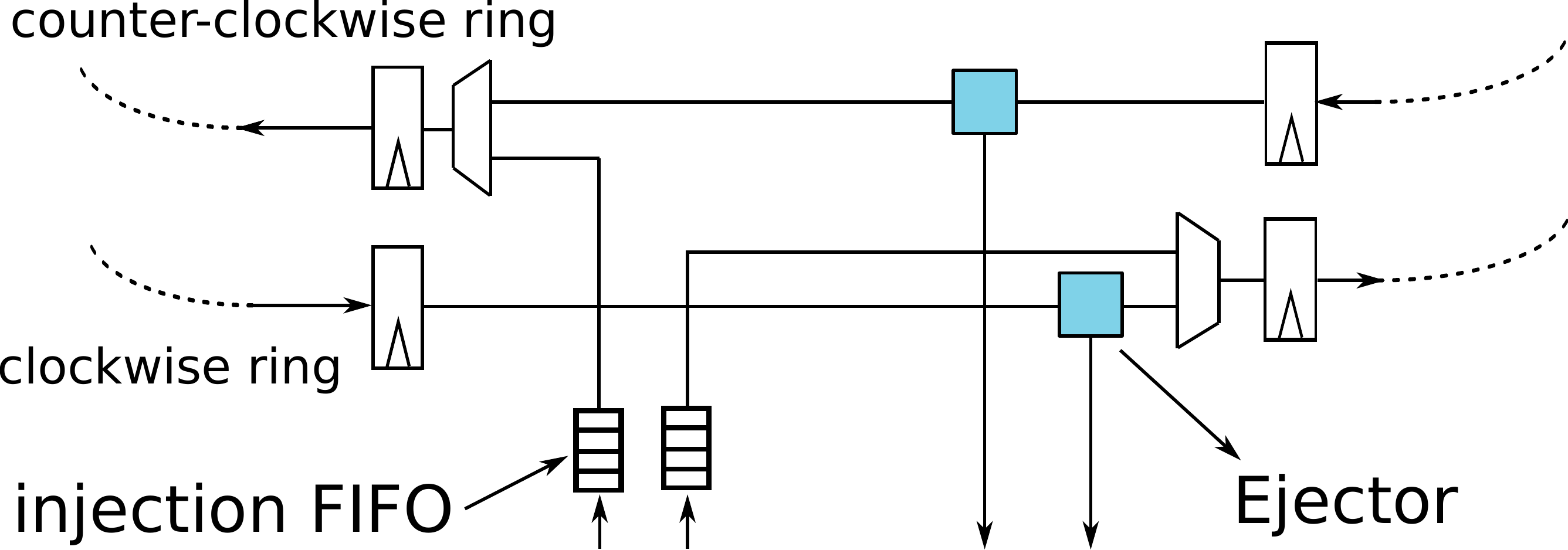}
\caption{\footnotesize \bf Node router.}
\label{fig:node-router}
\vspace{-0.1in}
\end{figure}

At each node on a local ring, we place a single node router, shown in
Figure~\ref{fig:node-router}. A node router is very simple: it passes
through circulating traffic, allows new traffic to enter the ring
through a MUX, and allows traffic to leave the ring when it arrives at
its destination. Each router contains one pipeline register for the
router stage, and one pipeline register for link traversal, so the
router latency is exactly one cycle and the per-hop latency is two
cycles. Such a design is very common in ring-based and
ring-like designs (e.g.,~\cite{kim09}).

As flits enter the router on the ring, they first travel to the
ejector. Because we use bidirectional rings, each node router has two
ejectors, one per direction.\footnote{For simplicity, we assume that up
  to two ejected flits can be accepted by the processor or reassembly
  buffers in a single cycle. For a fair comparison, we also implement
  two-flit-per-cycle ejection in our baselines.} Note that
the flits constituting a packet may arrive out-of-order and at widely
separated times. Re-assembly into packets is thus necessary. Packets
are re-assembled and reassembly buffers are managed using the
Retransmit-Once scheme, borrowed from the CHIPPER bufferless router
design~\cite{chipper}. With this scheme, receivers reassemble packets
in-place in MSHRs (Miss-Status Handling Registers~\cite{kroft81}),
eliminating the need for separate reassembly buffers. The key idea in
Retransmit-Once is to avoid ejection backpressure-induced deadlocks by
ensuring that all arriving flits are consumed immediately at their
receiver nodes. When a flit from a new packet arrives, it allocates a
new reassembly buffer slot if available. If no slot is available, the
receiver drops the flit and sets a bit in a retransmit queue which
corresponds to the sender and transaction ID of the dropped
flit. Eventually, when a buffer slot becomes available at the
receiver, the receiver reserves the slot for a sender/transaction ID
in its retransmit queue and requests a retransmit from the
sender. Thus, all traffic arriving at a node is consumed (or dropped)
immediately, so ejection never places backpressure on the
ring. Retransmit-Once hence avoids protocol-level
deadlock~\cite{chipper}. Furthermore, it ensures that a ring full of
flits always drains, thus ensuring forward progress (as we will
describe more fully in~\cite{hird}).

After locally-destined traffic is removed from the ring, the remaining
traffic travels to the injection stage. At this stage, the router
looks for ``empty slots,'' or cycles where no flit is present on the
ring, and injects new flits into the ring whenever they are queued for
injection. The injector is even simpler than the ejector, because it
only needs to find cycles where no flit is present and insert new
flits in these slots. Note that we implement two separate injection
buffers (FIFOs), one per ring direction; thus, two flits can be
injected into the network in a single cycle. A flit enqueues for
injection in the direction that yields a shorter traversal toward its
destination.



\subsection{Bridge Routers}
\label{sec:bridge_router}

\begin{figure}[h]
\centering
\vspace{-0.15in}
\includegraphics[width=2.8in]{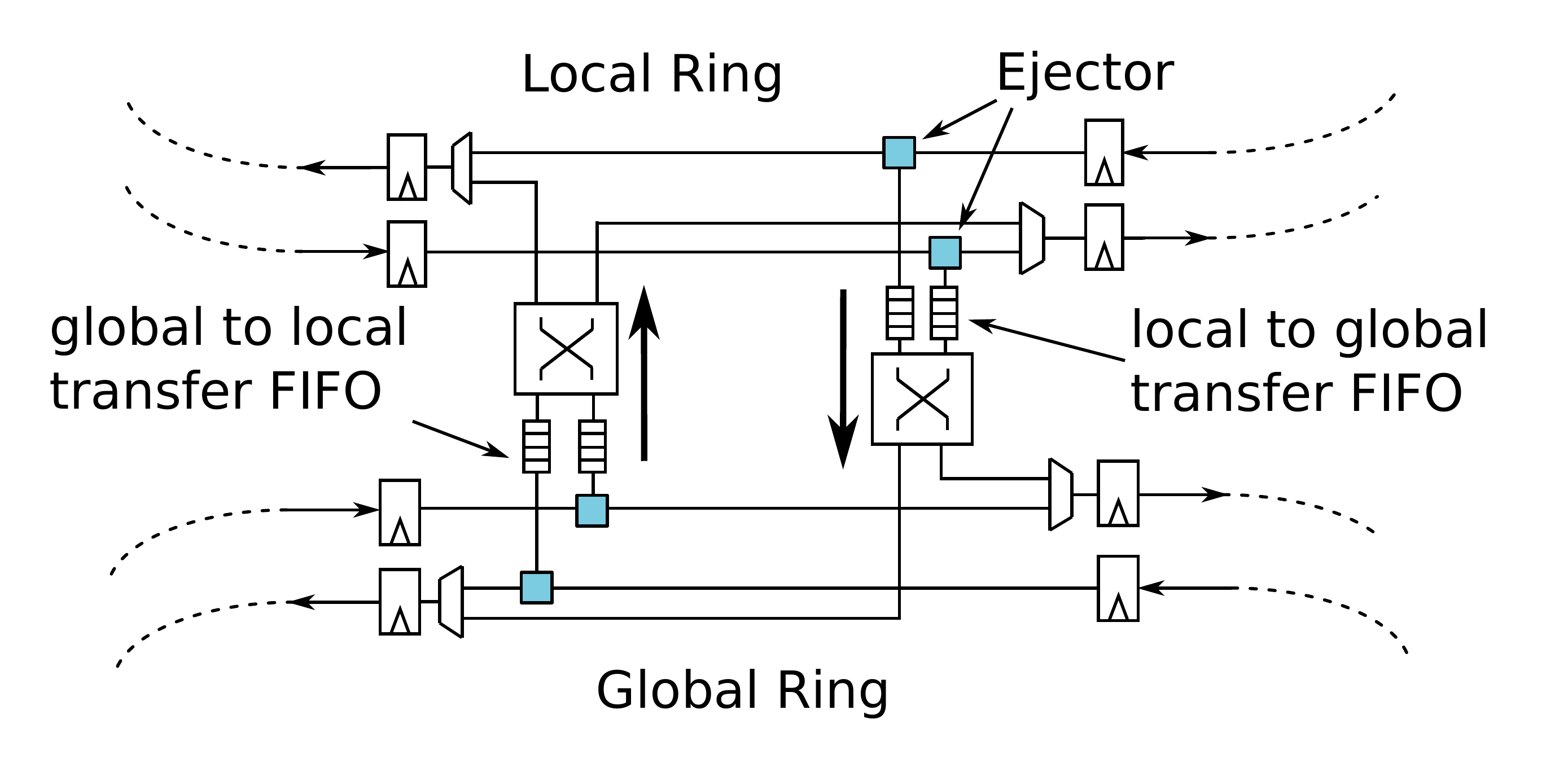}
\vspace{-0.15in}
\caption{\footnotesize \bf Bridge router.}
\label{fig:bridge-router}
\vspace{-0.1in}
\end{figure}

The \emph{bridge routers} connect a local ring and a global
ring, or a global ring with a higher-level global ring (if there are
more than two levels of hierarchy).  A high-level block diagram of a
bridge router is shown in Figure~\ref{fig:bridge-router}.  A bridge
router resembles two node routers, one on each of two rings, connected
by FIFO buffers in both directions.  When a flit arrives on one ring
that requires a transfer to the other ring (according to the routing
function described below in \S\ref{sec:routing}), it can leave its
current ring and wait in a FIFO as long as there is space
available. These \emph{transfer FIFOs} exist so that a transferring
flit's arrival need not be perfectly aligned with a free slot on the
destination ring.  However, this transfer FIFO will sometimes fill. In
that case, if any flit arrives that requires a transfer, the bridge
router simply does not remove the flit from its current ring; the flit
will continue to travel around the ring, and will eventually come back
to the bridge router, at which point there may be an open slot
available in the transfer FIFO. This is analogous to a
\emph{deflection} in hot-potato routing~\cite{hotpotato}, also known
as deflection routing, and has been used in recent on-chip mesh interconnect
designs to resolve
contention~\cite{casebufferless,chipper,minbd,glsvlsi,hotnets2010,sigcomm12}.  Note that to
ensure that flits are \emph{eventually} delivered, despite any
deflections that may occur, we introduce two \emph{guarantee
mechanisms} in~\cite{hird}.  Finally, note that
deflections may cause flits to arrive out-of-order (this is
fundamental to any non-minimal adaptively-routed network). Because we
use Retransmit-Once~\cite{chipper}, packet reassembly works despite
out-of-order arrival.

The bridge router uses \emph{crossbars} to allow a flit ejecting from either
ring direction in a bidirectional ring to enqueue for injection in either
direction in the adjoining ring. When a flit transfers, it picks the ring
direction that gives a shorter distance, as in a node router. However, these
crossbars actually allow for a more general case: the bridge router can
actually join several rings together by using larger crossbars. For our network
topology, we use hierarchical rings. We use wider global rings than local rings
(analogous to a \emph{fat tree}~\cite{cm}) for performance reasons. These wider rings
perform logically as separate rings as wide as one flit. Although not shown in
the figure for simplicity, the bridge router in such a case uses a larger
crossbar and has one ring interface (including transfer FIFO) per ring-lane in
the wide global ring. The bridge router then load-balances flits between rings
when multiple lanes are available. (The crossbar and transfer FIFOs are fully
modeled in our evaluations.)

\begin{figure*}[h]
\centering
\vspace{-0.4in}
\subfloat[4-, 8-, and 16-bridge hierarchical ring designs.]{
\includegraphics[width=2.7in]{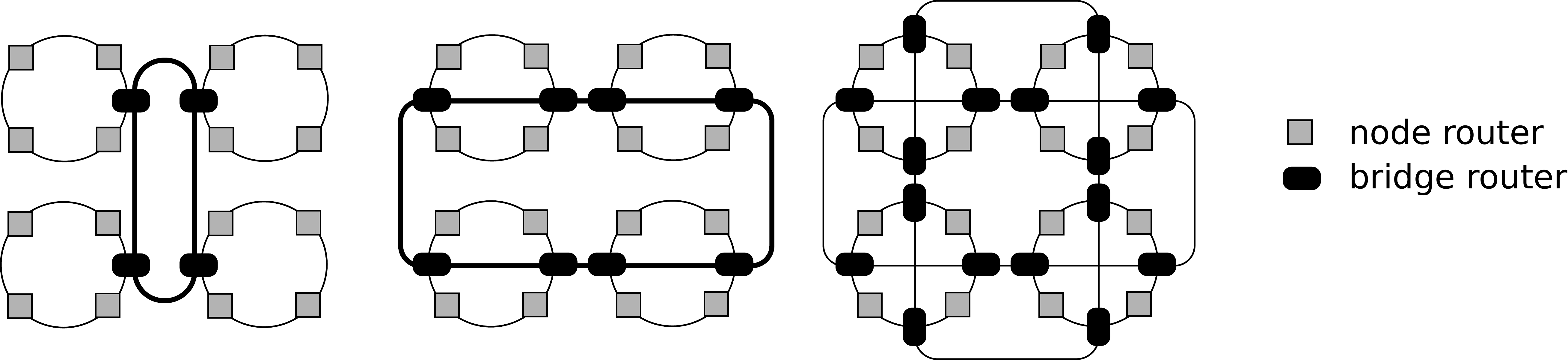}
\label{fig:topology}
}
\subfloat[Three-level hierarchy (8x8).]{
\includegraphics[width=1.5in]{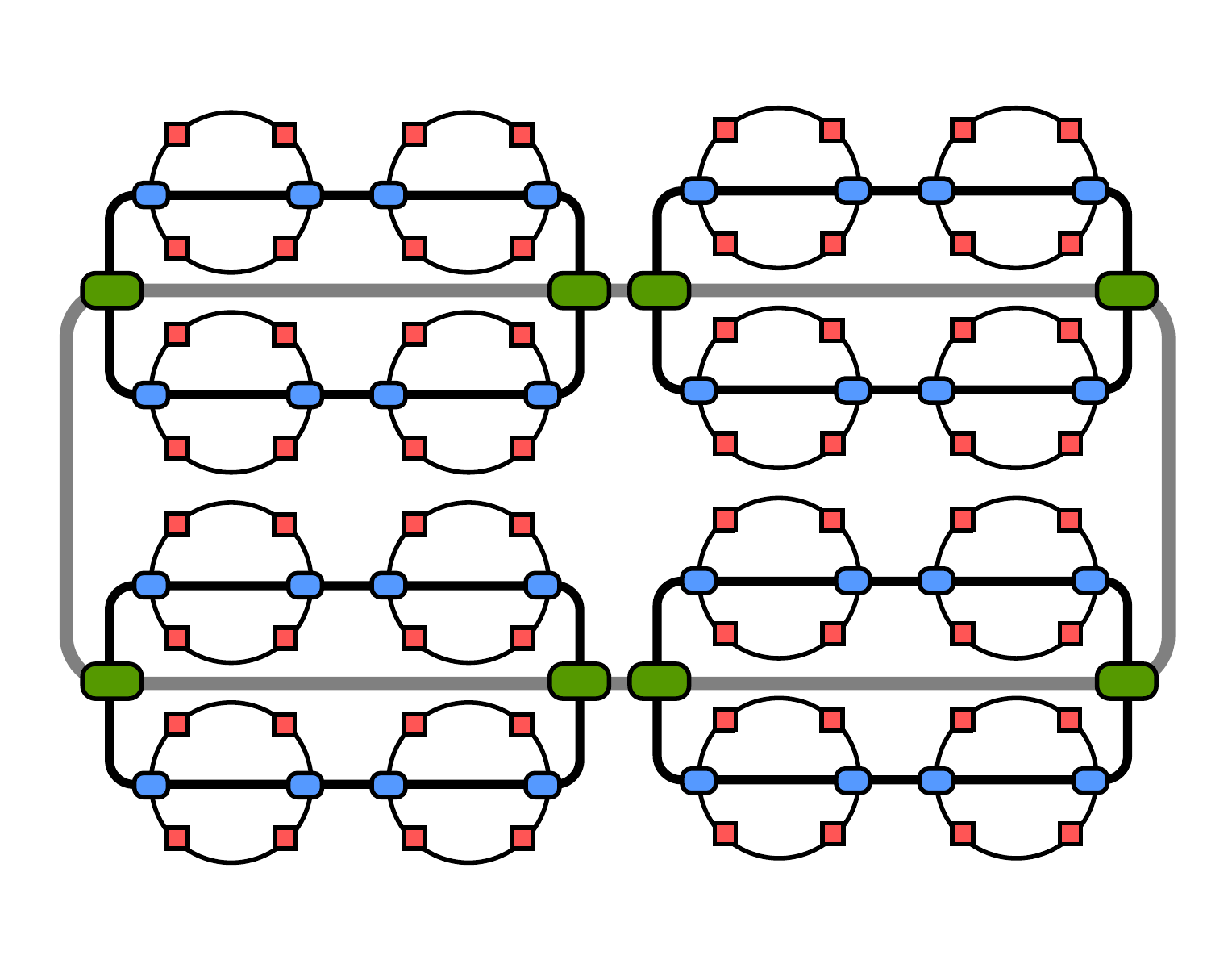}
\label{fig:scale}
}
\caption{\footnotesize \bf Hierarchical ring design of HiRD.}
\end{figure*}

When building a two-level design, there are many different arrangements of
global rings and bridge routers that can efficiently link the local rings
together. Figure~\ref{fig:topology} shows three designs denoted by the number
of bridge routers in total: 4-bridge, 8-bridge, and 16-bridge. We assume an
8-bridge design for the remainder of this paper. 
Also, note that the hierarchical structure that we
propose can be extended to more than two levels. We use a 3-level hierarchy,
illustrated in Figure~\ref{fig:scale}, to build a 64-node network.

Finally, in order to address a potential deadlock case (which will be
explained more in~\cite{hird}), bridge routers
implement a special \emph{Swap Rule}. The Swap Rule states that when
the flit that just arrived on each ring requires a transfer to the
other ring, the flits can be \emph{swapped}, bypassing the transfer
FIFOs altogether. This requires a bypass datapath (which is fully
modeled in our hardware evaluations). It ensures correct operation in
the case when transfer FIFOs in both directions are full. Only one
swap needs to occur in any given cycle, even when the bridge router
connects to a wide global ring. Note that because the swap rule
requires this bypass path, the behavior is always active (it would be
more difficult to definitively identify a deadlock and enable the
behavior only in that special case). The Swap Rule may cause flits to
arrive out-of-order when some are bypassed in this way, but the
network already delivers flits out-of-order, so correctness is not
compromised.

\ignore{
\floatname{algorithm}{Ruleset}
\begin{algorithm}[h]
\caption{Bridge Router Operation}
\label{rule:bridge-rule}
\footnotesize{
\begin{algorithmic}
\STATE 1. \textbf{Local Pass-through:} Flits arriving on the local ring
that are addressed to a destination on that local ring
remain on the ring.
\STATE 2. \textbf{Global Pass-through:} Flits arriving on the global
ring interface that are addressed to a destination not on this bridge
router's local ring remain on the global ring.
\STATE 3. \textbf{Local Ejection:} Flits arriving on the local ring
that are addressed to a destination on another local ring are ejected
from the local ring and placed into the global ring injection FIFO, if
it is not full. If the FIFO is full, the flit is not transferred, but
remains on the local ring (a \emph{deflection}).
\STATE 4. \textbf{Global Injection:} When a flit is present in the
global ring injection FIFO, and no flit arrives on the global ring
interface, the flit waiting to inject may be injected.
\STATE 5. \textbf{Global Ejection:} Flits arriving on the global ring
that are addressed to a destination on the local ring are ejected from
the global ring and placed into the local ring injection FIFO, if it
is not full. If the FIFO is full, the flit is not transferred but
remains on the global ring (a \emph{deflection}).
\STATE 6. \textbf{Local Injection:} When a flit is present in the
local ring injection FIFO, and no flit arrives on the local ring
interface, the flit waiting to inject may be injected.
\STATE 7. \textbf{Dual-Transfer Swap:} As a special case, when a flit
on the local ring and a flit on the global ring wish to eject
simultaneously, they may do so (i.e., they are swapped), which enables
the creation of empty slots for other flits to be injected. This rule
eliminates full-buffer deadlock, as it is performed even if both FIFOs
are full.
\end{algorithmic}
}
\end{algorithm}
\floatname{algorithm}{Algorithm}
}

\ignore{
Of particular note is the \emph{Dual-Transfer Swap} rule. This rule
allows flits from the global and local ring to swap places without
``temporary space,'' or empty FIFO slots, available. A swap flit
transfer occurs whenever the flit present on each side of a bridge
router requires a transfer. This dual-transfer swap is necessary to
ensure forward progress and avoid deadlock when the transfer FIFOs
become full, as we describe in the next section. However, we allow a
dual-transfer swap to occur even when the FIFO buffers are not full,
since its associated logic and bypass paths are already present for
deadlock-avoidance reasons.

}

\subsection{Routing}
\label{sec:routing}

Finally, we briefly address routing. Because a hierarchical ring design is
fundamentally a \emph{tree}, routing is very simple: when a flit is destined
for a node in another part of the hierarchy, it first travels \emph{up} the
tree (to more global levels) until it reaches a common ancestor of its source
and its destination, and then it travels \emph{down} the tree to its
destination. Concretely, each node's address can be written as a series of
parts, or digits, corresponding to each level of the hierarchy (these trivially
could be bitfields in a node ID). A ring can be identified by the common prefix
of all routers on that ring; the root global ring has a null (empty) prefix,
and local rings have prefixes consisting of all digits but the last one. If a
flit's destination does not match the prefix of the ring it is on, it takes any
bridge router to a more global ring. If a flit's destination does match the
prefix of the ring it is on (meaning that it is traveling down to more
local levels), it takes any bridge router which connects to the next level, until
it finally reaches the local ring of its destination and ejects at the node
with a full address match.


\section{Guaranteed Delivery: Correctness in Hierarchical Ring Interconnects}
\label{sec:livelock}
\label{sec:guarantees}

In order for the system to operate correctly, the interconnect must guarantee
that every flit is eventually delivered to its destination. HiRD ensures
correct operation through two mechanisms that provide two guarantees: the
\emph{injection guarantee} and the \emph{transfer guarantee}. The injection
guarantee ensures that any flit waiting to inject into a ring will eventually
be able to enter that ring. The transfer guarantee ensures that any flit
waiting to enter a bridge router's transfer queue will eventually be granted a
slot in that queue.

\begin{figure}[h]
\centering
\vspace{-0.2in}
\includegraphics[width=2.8in]{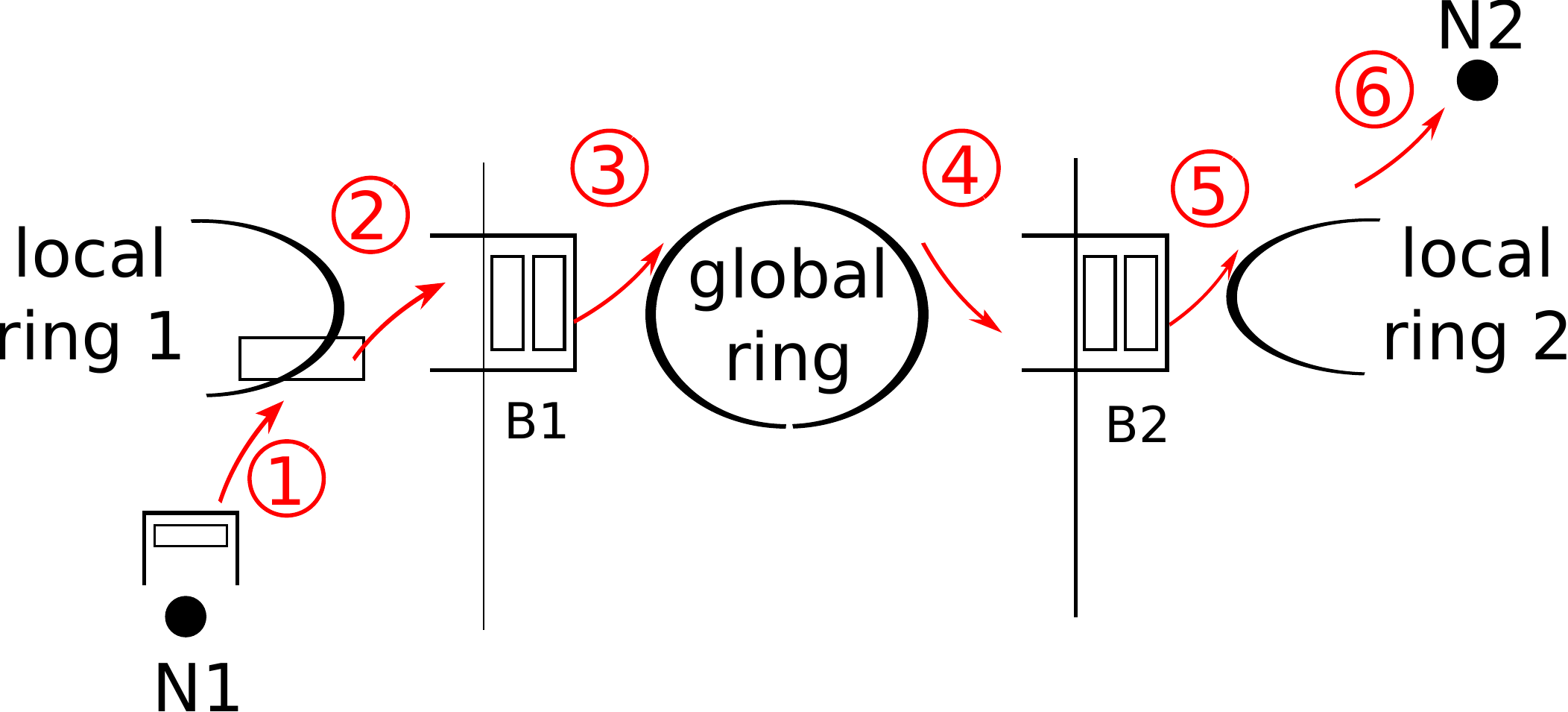}
\caption{\footnotesize \bf The need for the injection and transfer guarantees:
  contention experienced by a flit during its journey.}
\label{fig:guarantee-motivation}
\vspace{-0.15in}
\end{figure}

To understand the need for each guarantee, let us consider an example, shown in
Figure~\ref{fig:guarantee-motivation}. A flit is enqueued for network injection
at node N1 on the leftmost local ring. This flit is destined for node N2 on the
rightmost local ring; hence, it must traverse the leftmost local ring, then the
global ring in the center of the figure, followed by the rightmost local ring.
The flit transfers rings twice, at the two bridge routers B1 and B2 shown in
the figure. The figure also indicates the six points (labeled as \mycirc{1} to
\mycirc{6}) at which the flit moves from a queue to a ring or vice-versa: the
flit first enters N1's injection queue, transfers to the leftmost local ring
\mycirc{1}, the bridge router B1 \mycirc{2}, the global ring \mycirc{3}, the
bridge router B2 \mycirc{4}, the rightmost local ring \mycirc{5}, and finally
the destination node N2 \mycirc{6}.

In the worst case, when the network is heavily contended, the flit
could wait for an unbounded amount of time at \mycirc{1} to
\mycirc{5}. First, recall that to enter any ring, a flit must wait for an empty
slot on that ring (because the traffic on the ring continues along the
ring once it has entered, and thus has higher priority than any new
traffic). Because of this, the flit traveling from node N1 to N2 could
wait for an arbitrarily long time at \mycirc{1}, \mycirc{3}, and \mycirc{5}  if no other
mechanism intercedes. This first problem is one of \emph{injection
  starvation}, and we address it with the \emph{injection guarantee}
mechanism described below. Second, recall that a flit that needs to
transfer from one ring to another via a bridge router enters that
bridge router's queue, but if the bridge router's queue is full, then
the transferring flit must make another trip around its current ring
and try again when it next encounters a bridge router. Because of this
rule, the flit traveling from N1 to N2 could be \emph{deflected} an
arbitrarily large number of times at \mycirc{2} and \mycirc{4} (at entry to
bridge routers B1 and B2) if no other mechanism intercedes. This
second problem is one of \emph{transfer starvation}, and we address it
with the \emph{transfer guarantee} mechanism described below.

\emph{Our goal} in this section is to demonstrate that HiRD provides both the
injection guarantee (\S\ref{sec:guarantee-injection}) and the transfer guarantee
(\S\ref{sec:guarantee-transfer}) mechanisms. We show correctness
in \S\ref{sec:guarantee-correctness}, and quantitatively evaluate both
mechanisms in \S\ref{sec:eval-guarantees} and in~\cite{hird-safari-tr}.

\subsection{Preventing Injection Starvation: Injection Guarantee}
\label{sec:guarantee-injection}

The \emph{injection guarantee} ensures that every router on a ring can
eventually inject a flit. This guarantee is provided by a very simple
throttling-based mechanism: when any node is starved (cannot inject a flit)
past a threshold number of cycles, it asserts a signal to a global controller,
which then throttles injection from every other node. No new traffic will enter
the network while this throttling state is active. All existing flits in the
network will eventually drain, and the starved node will be able to finally
inject its new flit. At that time, the starved node de-asserts its throttling
request signal to the global controller, and the global controller subsequently
allows all other nodes to resume normal operation. 

Note that this injection guarantee can be implemented in a hierarchical manner
to improve scalability. In the hierarchical implementation, each individual
local ring in the network monitors only its own injection and throttles
injection locally if any node in it is starved. After a threshold number of
cycles.\footnote{In our evaluation, we set this threshold to be 100 cycles.} if
any node in the ring still cannot inject, the bridge routers connected to that
ring start sending throttling signals to any other ring in the next level of
the ring hierarchy they are connected to. In the worst case, every local ring
stops accepting flits and all the flits in the network drain and eliminate any
potential livelock or deadlock. Designing the delivery guarantee this way 
requires two wires in each ring and small design overhead at
the bridge router to propagate the throttling signal across hierarchy levels. In our
evaluation, we faithfully model this hierarchical design.

\ignore{
Note that a global throttling-based approach is not necessarily the
most optimal method to ensure a flit has an empty slot to enter the
network. In the best case, exactly one other new injection could be
throttled in order to create one empty slot into which the starved
node could inject its flit (i.e., a one-for-one trade in new flit
injections). \emph{However}, such a scheme would require significant
global coordination and run-time planning: some central authority
would have to find an empty slot in the network (in the worst case,
when the network is completely full everywhere, wait for some flit to
be delivered and removed from its ring) and explicitly mark that slot
as reserved for the starved node. In contrast, as long as starvation
events are \emph{rare}, it is much simpler to temporarily inhibit
injection from all nodes until the problem is (quickly) addressed as
space is freed. In fact, in our evaluations of real application works,
we observe that injection starvation events are exceedingly rare and
hence their bandwidth efficiency has little impact on overall
throughput. Our quantitative evaluation of the injection guarantee
mechanism in \S\ref{sec:eval-guarantees} will show that sufficient
throughput remains even in a worst-case synthetic traffic pattern for
injection starvation.
}

\subsection{Ensuring Ring Transfers: Transfer Guarantee}
\label{sec:guarantee-transfer}

The \emph{transfer guarantee} ensures that any flit waiting to transfer from
its current ring to another ring via a bridge router will eventually be able to
enter that bridge router's queue. Such a guarantee is non-trivial because the
bridge router's queue is finite, and when the destination ring is congested, a
slot may become available in the queue only infrequently. In the worst case, a
flit in one ring may circulate indefinitely, finding a bridge router to its
destination ring with a completely full queue each time it arrives at the
bridge router. The transfer guarantee ensures that any such circulating flit
will eventually be granted an open slot in the bridge router's transfer queue.
Note in particular that this guarantee is \emph{separate from} the injection
guarantee: while the injection guarantee ensures that the bridge router will be
able to inject flits from its transfer queue into the destination ring (and
hence, have open slots in its transfer queue eventually), these open transfer
slots may not be distributed \emph{fairly} to flits circulating on a ring
waiting to transfer through the bridge router. In other words, some flit may
always be ``unlucky'' and never enter the bridge router if slots open at the
wrong time. The transfer guarantee addresses this problem.

In order to ensure that any flit waiting to transfer out of a ring
eventually enters its required bridge router, each bridge router
\emph{observes a particular slot on its source ring} and monitors for
flits that are ``stuck'' for more than a threshold number of
retries. (To observe one ``slot,'' the bridge router simply examines
the flit in its ring pipeline register once every N cycles, where N is
the latency for a flit to travel around the ring once.) If any flit
circulates in its ring more than this threshold number of times, the
bridge router reserves the next open available entry in its transfer queue for
this flit (in other words, it will refuse to accept other flits for
transfer until the ``stuck'' flit enters the queue). Because of the
injection guarantee, the head of the transfer queue must inject into
the destination ring eventually, hence an entry must become available eventually,
and the stuck flit will then take the entry in the transfer queue the
next time it arrives at the bridge router. Finally, the slot which the
bridge router observes rotates around its source ring: whenever the
bridge router observes a slot the second time, if the flit that
occupied the slot on first observation is no longer present (i.e.,
successfully transferred out of the ring or ejected at its
destination), then the bridge router begins to observe the \emph{next}
slot (the slot that arrives in the next cycle). In this way, every
slot in the ring is observed eventually, and any stuck flit will thus
eventually be granted a transfer.

\subsection{Putting it Together: Guaranteed Delivery}
\label{sec:guarantee-correctness}


Before we prove the correctness of these mechanisms in detail, it is helpful to
summarize the basic operation of the network once more. A flit can inject into
a ring whenever a free slot is present in the ring at the injecting router
(except when the injecting router is throttled by the injection guarantee
mechanism). A flit can eject at its destination whenever it arrives, and
destinations always consume flits as soon as they arrive (which is ensured
despite finite reassembly buffers using the Retransmit-Once mechanism~\cite{chipper}, as
already described in \S\ref{sec:mech-node-router}). A flit transfers between
rings via a transfer queue in a bridge router, first leaving its source ring to
wait in the queue and then injecting into its destination ring when at the head
of the queue, and can enter a transfer queue whenever there is a free entry in
that transfer queue (except when the entry is reserved for another flit by the
transfer guarantee mechanism). Finally, when two flits at opposite ends of a
bridge router each desire to to transfer through the bridge router, the
\emph{Swap Rule} allows these flits to exchange places directly, bypassing the
queues (and ensuring forward progress).

Our proof is structured as follows: we first argue that if no new flits enter
the network, then the network will drain in finite time. The injection
guarantee ensures that any flit can enter the network. Then, using the
injection guarantee, transfer guarantee, the swap rule, and the fact that the
network is hierarchical, any flit in the network can eventually reach any ring
in the network (and hence, its final destination ring). Because all flits in a
ring continue to circulate that ring, and any node on a ring must consume any
flits that are destined for that node, final delivery is ensured once a flit
reaches its final destination ring.

\noindent\textbf{Network drains in finite time.} Assume no new flits enter the
network (for now). A flit could only be stuck in the network indefinitely if
transferring flits create a cyclic dependence between completely full rings.
Otherwise, if there are no dependence cycles, then if one ring is full and
cannot accept new flits because other rings will not accept \emph{its} flits,
then eventually there must be some ring which depends on no other ring (e.g., a
local ring with all locally-destined flits), and this ring will drain first,
followed by the others feeding into it. However, because the network is
hierarchical (i.e., a tree), the only cyclic dependences possible are between
rings that are immediate parent and child (e.g., global ring and local ring, in
a two-level hierarchy). The \emph{Swap Rule} ensures that when a parent and
child ring are each full of flits that require transfer to the other ring, then
transfer is always possible, and forward progress will be ensured. Note in
particular that we do not require the injection or transfer guarantee for the
network to drain. Only the \emph{Swap Rule} is necessary to ensure that no deadlock
will occur. \ignore{As long as flits continue to drain, no flit waiting in a
transfer queue can be starved from injecting into its destination ring for more
than a finite amount of time.}

\noindent\textbf{Any node can inject.} Now that we have shown that the network
will drain if no new flits are injected, it is easy to see that the injection
guarantee ensures that any node can eventually inject a flit: if any node is
starved, then all nodes are throttled, no new flit enters the network, and the
network must eventually drain (as we just showed), at which point the starved
node will encounter a completely empty network into which to inject its flit.
(It likely will be able to inject before the network is completely empty, but
in the worst case, the guarantee is ensured in this way.)

\noindent\textbf{All flits can transfer rings and reach their destination
rings.} With the injection guarantee in place, the transfer guarantee can be
shown to provide its stated guarantee as follows: because of the injection
guarantee, a transfer queue in a bridge router will always inject its head flit
in finite time, hence will have an open entry to accept a new transferring flit
in finite time. All that is necessary to ensure that \emph{all} transferring
flits eventually succeed in their transfers is that \emph{any} flit stuck for
long enough gets an available entry in the transfer queue. The transfer
guarantee does exactly this by observing ring slots in sequence and reserving a
transfer queue entry when a flit becomes stuck in a ring. Because the mechanism
will eventually observe every slot in the ring, all flits will be allowed to make their
transfers eventually. Hence, all flits can continue to transfer rings until
reaching their destination rings (and thus, their final destinations).

\subsection{Hardware Cost}

Our injection and transfer guarantee mechanisms have low hardware overhead. To implement the
injection guarantee, one counter is required for each injection point.
This counter tracks how many cycles have elapsed while injection is starved,
and is reset whenever a flit is successfully injected. Routers communicate with
the throttling arbitration logic with only two wires, one to signal blocked
injection and one control line that throttles the router. The wiring is done
hierarchically instead of globally to minimize the wiring cost (\S\ref{sec:guarantee-injection}). Because the
correctness of the algorithm does not depend on the delay of these wires, and
the injection guarantee mechanism is activated only rarely (in fact, \emph{never} for
our evaluated realistic workloads), the signaling and central coordinator need not be
optimized for speed. To provide the transfer guarantee, each bridge router
implements ``observer'' functionality for each of the two rings it sits on, and
the observer consists only of three small counters (to track the current
timeslot being observed, the current timeslot at the ring pipeline register in
this router, and the number of times the observed flit has circled the ring)
and a small amount of control logic.  Importantly, note that neither mechanism
impacts the router critical path nor affects the router datapath (which
dominates energy and area).



\section{Evaluation Methodology}
\label{sec:meth}


We perform our evaluations using a cycle-accurate simulator of a CMP system
with 1.6GHz interconnect to provide application-level performance
results~\cite{NOCulator}. Our simulator is publicly available and includes
the source code of all mechanisms we evaluated~\cite{NOCulator}.
Tables~\ref{table:sysparams}~and~\ref{table:topoparams} provide the configuration
parameters of our simulated systems.

\linespread{1}
\begin{table*}[h]
\centering
\footnotesize{
\begin{tabular}{|l||p{9.3cm}|}
\hline
\textbf{Parameter} & \textbf{Setting} \\
\hline
\hline
System topology & CPU core and shared cache slice at every node \\
\hline
Core model & Out-of-order, 128-entry ROB, 16 MSHRs (maximum simultaneous outstanding requests) \\
\hline
Private L1 cache & 64 KB, 4-way associative, 32-byte block size \\
\hline
Shared L2 cache & Perfect (always hits) to stress the network and
penalize our reduced-capacity deflection-based design; cache-block-interleaved
 \\
\hline
Cache coherence & Directory-based protocol (based on SGI
Origin~\cite{laudon97}), directory entries co-located with
shared cache blocks \\
\hline
Simulation length & 5M-instruction warm-up, 25M-instruction active execution per node
~\cite{casebufferless,chipper,hat-sbac-pad,minbd} \\
\hline
\end{tabular}
}
\caption{\footnotesize \bf Simulation and system configuration parameters.}
\vspace{-0.15in}
\label{table:sysparams}
\end{table*}

\begin{table*}
\centering
\footnotesize{
\begin{tabular}{|l|l|p{6.5cm}|}
\hline
Parameter & Network & Setting \\
\hline
\hline
\multirow{4}{*}{Interconnect Links} & Single Ring & Bidirectional, \textbf{4x4:} 64-bit and 128-bit width, \textbf{8x8:} 128-bit and 256-bit width \\
\cline{2-3}
& Buffered HRing & Bidirectional, \textbf{4x4:} 3-cycle per-hop latency (link+router); 64-bit local and 128-bit global rings, \textbf{8x8:} three-level hierarchy, 4x4 parameters, with second-level rings connected by a 256-bit third-level ring \\
\cline{2-3}
& HiRD & \textbf{4x4:} 2-cycle (local), 3-cycle (global) per-hop
latency (link+router); 64-bit local ring, 128-bit global ring; \textbf{8x8:} 4x4
parameters, with second-level rings connected by a 256-bit third-level
ring \\
\hline 
\multirow{6}{*}{Router} & Single Ring & 1-cycle per-hop latency (as in~\cite{kim09nocarc}) \\
\cline{2-3}
& Buffered HRing & Node (NIC) and bridge (IRI) routers based
on~\cite{ravindran97}; 4-flit in-ring and transfer
FIFOs. Bidirectional links of dual-flit width (for fair comparison
with our design). Bubble flow control~\cite{bubbleflow} for deadlock
freedom. \\
\cline{2-3}
& HiRD & Local-to-global buffer depth of 1, global-to-local buffer depth of 4 \\
\hline
\end{tabular}
}
\caption{\footnotesize \bf Network parameters.}
\vspace{-0.25in}
\label{table:topoparams}
\end{table*}

Our methodology ensures a rigorous and isolated evaluation
of NoC capacity for especially cache-resident workloads, and has also been used in
other studies~\cite{casebufferless,chipper,hotnets2010,sigcomm12,minbd}. Instruction traces for the
simulator are taken using a Pintool~\cite{pin} on representative
portions of SPEC CPU2006 workloads.

We mainly compare to a single
bidirectional ring and a state-of-the-art buffered hierarchical
ring~\cite{ravindran97}. 
Also, note that while there are many possible ways to
optimize each baseline (such as congestion control~\cite{hat-sbac-pad,hotnets2010,sigcomm12}, adaptive routing
schemes, and careful parameter tuning), we assume a fairly typical
aggressive configuration for each.

\ignore{
Note in particular that the buffered hierarchical ring baseline, which
is the most closely-related design to HiRD, is based on the design
evaluated by Ravindran et al.~\cite{ravindran97}. The key difference
between this buffered hierarchical ring design and HiRD is that
because all routers in the prior work have in-ring buffers, the
network performs token-based flow control with backpressure rather
than deflection-based flow control only at bridge routers as we
propose. The flow control is based on bubble flow
control~\cite{bubbleflow} to ensure deadlock freedom. As a result of
the more complex flow control, the routers in this prior work have
higher control logic complexity than the node routers in our design.
}


\noindent\textbf{Data Mapping.} We map data in a cache-block-interleaved way to
different shared L2 cache slices. This mapping is agnostic to the underlying locality.
As a result, it does not exploit the low-latency data access in the local ring.
One can design systematically better mapping in order to keep frequently used
data in the local ring as in~\cite{CloudCache,ccraik-safari-tr}. However, such a mapping mechanism is orthogonal
to our proposal and can be applied in all ring-based network designs.

\label{sec:meth-workloads}

\noindent\textbf{Application \& Synthetic Workloads.} The system is run with a
set of 60 multiprogrammed workloads. Each workload consists of one
single-threaded instance of a SPEC CPU2006 benchmark on each core, for a total
of either 16 (4x4) or 64 (8x8) benchmark instances per workload.
Multiprogrammed workloads such as these are representative of many common
workloads for large CMPs.  Workloads are constructed at varying network
intensities as follows: first, benchmarks are split into three classes (Low,
Medium and High) by L1 cache miss intensity (which correlates directly with
network injection rate), such that benchmarks with less than 5 misses per
thousand instructions (MPKI) are ``Low-intensity,'' between 5 and 50 are
``Medium-intensity,'' and above 50 MPKI are ``High-intensity.''  Workloads are
then constructed by randomly selecting a certain number of benchmarks from each
category. We form workload sets with four intensity mixes: High (H), Medium (M),
Medium-Low (ML), and Low (L), with 15 workloads in each (the average network injection
rates for each category are 0.47, 0.32, 0.18, and 0.03 flits/node/cycle,
respectively). 

\noindent\textbf{Multithreaded Workloads.} We use the GraphChi implementation
of the GraphLab framework~\cite{graphchi,graphlab}. The implementation we use
is designed to run efficiently on multi-core systems. The workload consists of
Twitter Community Detection (CD), Twitter Page Rank (PR), Twitter Connected
Components (CC), Twitter Triangle Counting (TC)~\cite{twitter_rv}, and Graph500
Breadth First Search (BFS). We simulated the representative portion of each
workload and each workload has a working set size of greater than 151.3$\,$MB.
On every simulation of these multithreaded workloads, we warm up the cache with
the first 5 million instructions, then we run the remaining code of the
representative portion.  



\ignore{
While some other works use multithreaded workloads, there is little
fundamental difference at the network level between coherence traffic
and traffic from threads with independent working sets. In addition,
synthetic traffic shows fundamental enhancements in network
performance regardless of workload.
}



\noindent\textbf{Energy \& Area.} We measure the energy and area of routers and links by
individually modeling the crossbar, pipeline registers, buffers, control logic,
and other datapath components. For links, buffers and datapath elements, we use
DSENT 0.91~\cite{dsent}. Control logic is modeled in Verilog RTL\ignore{synthesized with a
commercial design library}. Both energy and area are calculated based on a 45nm
technology. \green{The link lengths we assume are based on the floorplan of our designs,
which we describe in the next paragraph.}

\begin{figure}[h!]


\centering
\includegraphics[width=3.5in]{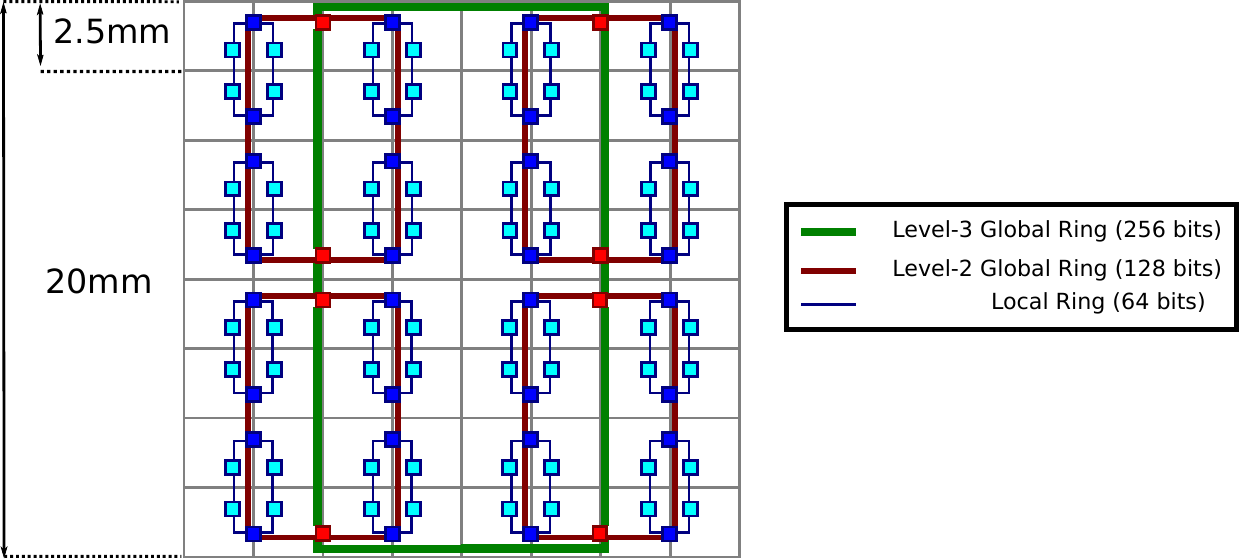}
\caption{\footnotesize \bf Assumed floorplan for HiRD 3-level (64-node)
  network. Two-level (16-node) network consists of one quadrant of
  this floorplan.}
\label{fig:hird-floorplan}
\end{figure}

\green{We assume the area of each core to be 2.5 mm $x$ 2.5 mm.} We assume a
2.5 mm link length for single-ring designs. For the hierarchical ring design,
we assume 1 mm links between local-ring routers, because the four routers on a
local ring can be placed at four corners that meet in a tiled design.
Global-ring links are assumed to be 5.0 mm \green{(i.e., five times as long as local links)}, because they span across two tiles
on average if local rings are placed in the center of each four-tile quadrant.
Third-level global ring links are assumed to be 10mm \green{(i.e., ten times as long as local links)} in the 8x8 evaluations.
This floorplan is illustrated in more detail in
Figure~\ref{fig:hird-floorplan} \green{for the 3-level (64-node) HiRD network. Note that one
quadrant of the floorplan of Figure~\ref{fig:hird-floorplan} corresponds to the floorplan of the 2-level (16-node) HiRD network.
We faithfully take into account all link lengths in our energy and area estimates for all designs.}

%
%
%
%


\noindent\textbf{Application Evaluation Metrics.} For multiprogrammed workloads, we present
application performance results using the commonly-used Weighted
Speedup metric~\cite{weighted_speedup,harmonic_speedup}. We use the maximum slowdown
metric to measure unfairness~\cite{stc,atlas,tcm,vandierendonck,fst,eiman-isca11,eiman-micro09,mcp,aergia,a2c,asm-micro15,mise-hpca13,dash-taco16,sms,bliss,bliss-tpds}.
\ignore{We take an application's performance running on a baseline buffered network listed
in Table~\ref{table:topoparams} as a
common baseline when computing weighted speedup on all designs, in
order to allow for fair comparison across designs.}

\section{Evaluation}
\label{sec:evaluation}

We provide a comprehensive evaluation of our proposed mechanism against other
ring baselines.  Since our goal is to provide a better ring design, our main
comparisons are to ring networks. However, we also provide sensitivity analyses
and comparisons to other network designs as well.



\subsection{Ring-based Network Designs}
\label{sec:eval-hrings}

\noindent\textbf{Multiprogrammed workloads}

\begin{figure*}[h!]
\vspace{-0.1in}
\centering
\includegraphics[width=5in]{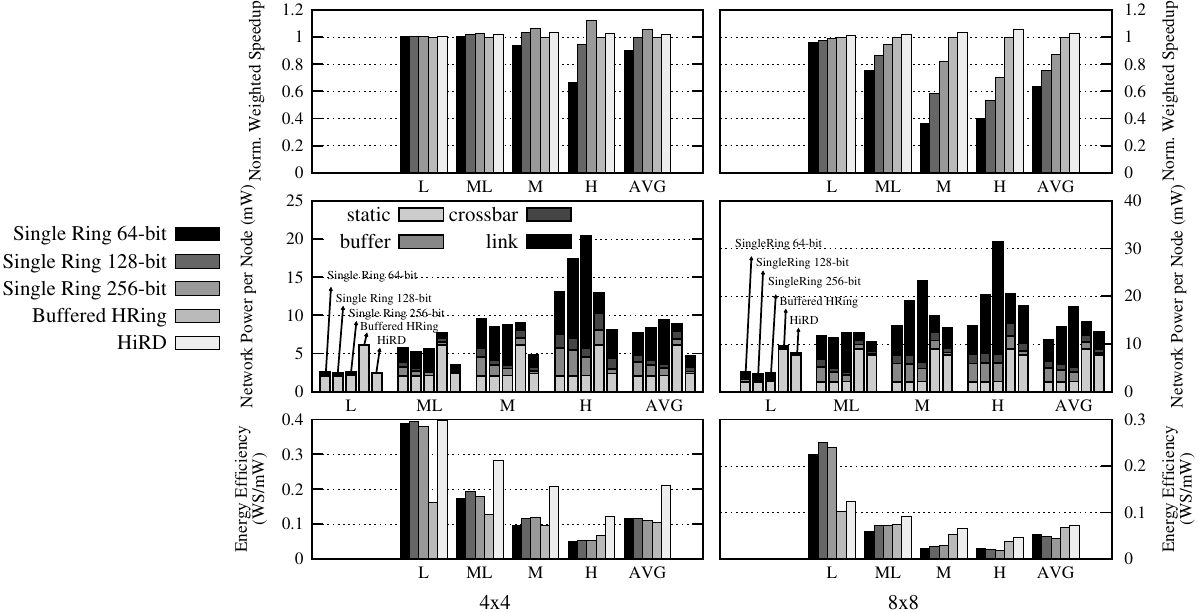}
\vspace{-0.05in}
\caption{\footnotesize \bf HiRD as compared to buffered hierarchical rings and a single-ring network.}
\label{fig:hird-buf-hring}
\end{figure*}

Figure~\ref{fig:hird-buf-hring} shows performance (weighted speedup
normalized per node), power (total network power normalized per node),
and energy-efficiency (perf./power) for 16-node and 64-node HiRD and
buffered hierarchical rings in~\cite{ravindran97}, using identical topologies, as well as a
single ring (with different bisection bandwidths).

1. A hierarchical topology yields significant performance advantages over a
single ring (i) when network load is high and/or (ii) when the network scales
to many nodes. As shown, the buffered hierarchical ring improves
performance by 7\% (and HiRD by 10\%) 
in high-load workloads at 16 nodes compared to a single ring with 128-bit links. The
hierarchical design also reduces power because hop count is reduced. Therefore,
link power reduces significantly with respect to a single ring.
\ignore{
2. HiRD improves performance over the buffered hierarchical ring in larger
networks, where the scaling advantage of hierarchy becomes more important.}
On average, in the 8x8 configuration, the buffered hierarchical ring network
obtains 15.6\% better application performance than the single ring with 256-bit
links, while HiRD attains 18.2\% higher performance.


2. Compared to the buffered hierarchical ring, HiRD has significantly
lower network power and better performance. On average, HiRD reduces total network power
(links and routers) by 46.5\% (4x4) and 14.7\% (8x8) relative to this
baseline. This reduction in turn yields significantly better energy
efficiency (lower energy consumption for buffers and slightly higher for links).\footnote{Note 
that the low intensity workloads in the 8x8 network is an exception.
HiRD reduces energy efficiency for these as the static link power becomes dominant
for them.} Overall, HiRD is the most energy-efficient of the
ring-based designs evaluated in this paper for both 4x4 and 8x8 network
sizes. HiRD also performs better than Buffered HRing due to the reasons explained
in the next section (\S\ref{sec:eval-synth}).


3. While scaling the link bandwidth increases the performance of a single ring
network, the network power increases 25.9\% when the link bandwidth increases
from 64-bit to 128-bit and  15.7\% when the link bandwidth increases from
128-bit to 256-bit because of higher dynamic energy due to wider links.
\ignore{Additionally, given the same amount of energy consumed by the network,
increasing link bandwidth increases the performance, which will increase the
power consumption as power is energy over time.} In addition, scaling the link
bandwidth is not a scalable solution as a single ring network performs worse
than the bufferred hierarchical ring baseline even when a 256-bit link is used.


We conclude that HiRD is effective in simplifying the design of the
hierarchical ring and making it more energy efficient, as we intended to as our design goal. We show that HiRD provides competitive performance compared
to the baseline buffered hierarchical ring design with equal or better energy
efficiency.



\noindent\textbf{Multithreaded workloads}

\begin{figure}
\vspace{-0.2in}
\centering
\includegraphics[width=5in]{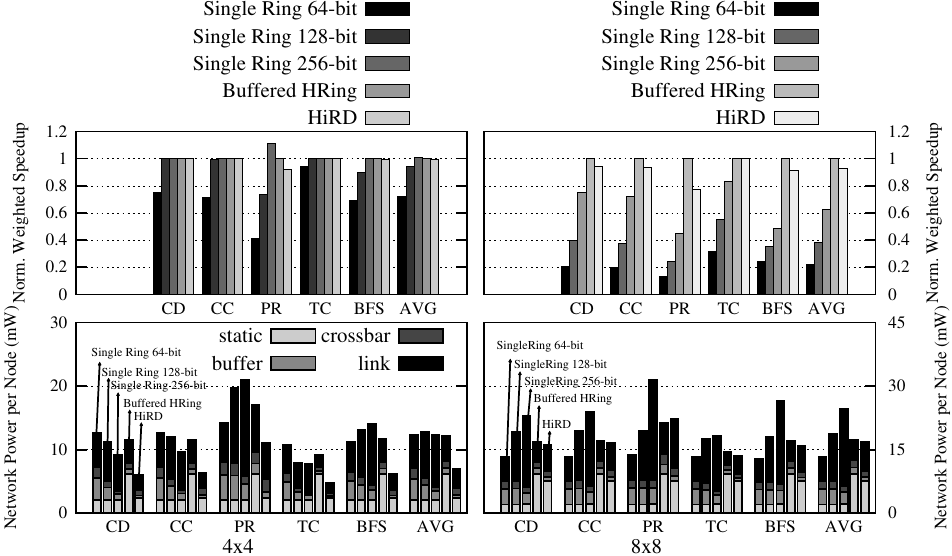}
\caption{\footnotesize \bf HiRD as compared to buffered hierarchical rings and a single-ring network on multithreaded workloads.}
\label{fig:hird-buf-hring-graph}
\end{figure}

\indent Figure~\ref{fig:hird-buf-hring-graph} shows the performance and power of HiRD on
multithreaded applications compared to a buffered hierarchical ring and a
single-ring network for both 16-node and 64-node systems. On average, HiRD
performs 0.1\% (4x4) and 0.73\% (8x8) worse than the buffered hierarchical ring.
However, on average, HiRD consumes 43.8\% (4x4) and 3.1\% (8x8) less power,
leading to higher energy efficiency. This large reduction in energy comes from the
elimination of most buffers in HiRD. 


Both the buffered hierarchical ring and HiRD outperform single ring networks,
and the performance improvement increases as we scale the size of the network.


Even though HiRD performs competitively with a buffered hierarchical ring
network in most cases, HiRD performs poorly on the Page Ranking application. We
observe that Page Ranking generates more non-local network traffic than other applications. As HiRD is
beneficial mainly at lowering the local-ring latency, it is unable to speed up
such non-local traffic, and is thus unable to help Page Ranking. In addition,
Page Ranking also has higher network traffic, causing more congestion in the
network (we observe 17.3\% higher average network latency for HiRD in an 8x8
network), and resulting in a performance drop for HiRD. However, it is
possible to use a different number of bridge routers as illustrated
in Figure~\ref{fig:topology}, to improve the performance of HiRD, which we will
analyze in Section~\ref{sec:eval-sensitivity}.  Additionally, it
is possible to apply a locality-aware cache mapping technique~\cite{CloudCache,ccraik-safari-tr}
in order to take advantage of lower local-ring latency in HiRD. 

We conclude that HiRD is effective in improving evergy efficiency significantly 
for both multiprogrammed and multithreaded applications.

\subsection{Synthetic-Traffic Network Behavior}
\label{sec:eval-synth}

\begin{figure}[h!]
\centering
\subfloat[Uniform Random, 4x4]{
\includegraphics[width=1.5in]{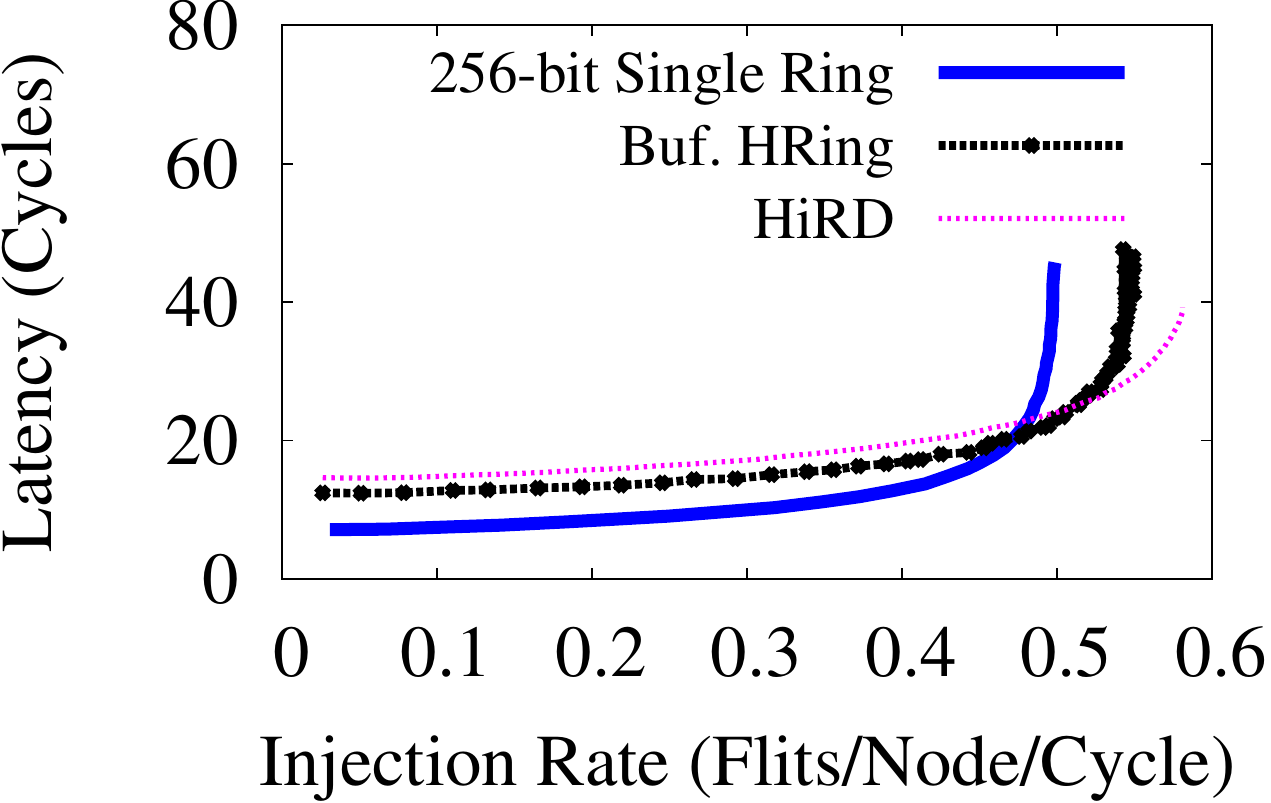}
\label{fig:synthUR}
}
\subfloat[Uniform Random, 8x8]{
\includegraphics[width=1.5in]{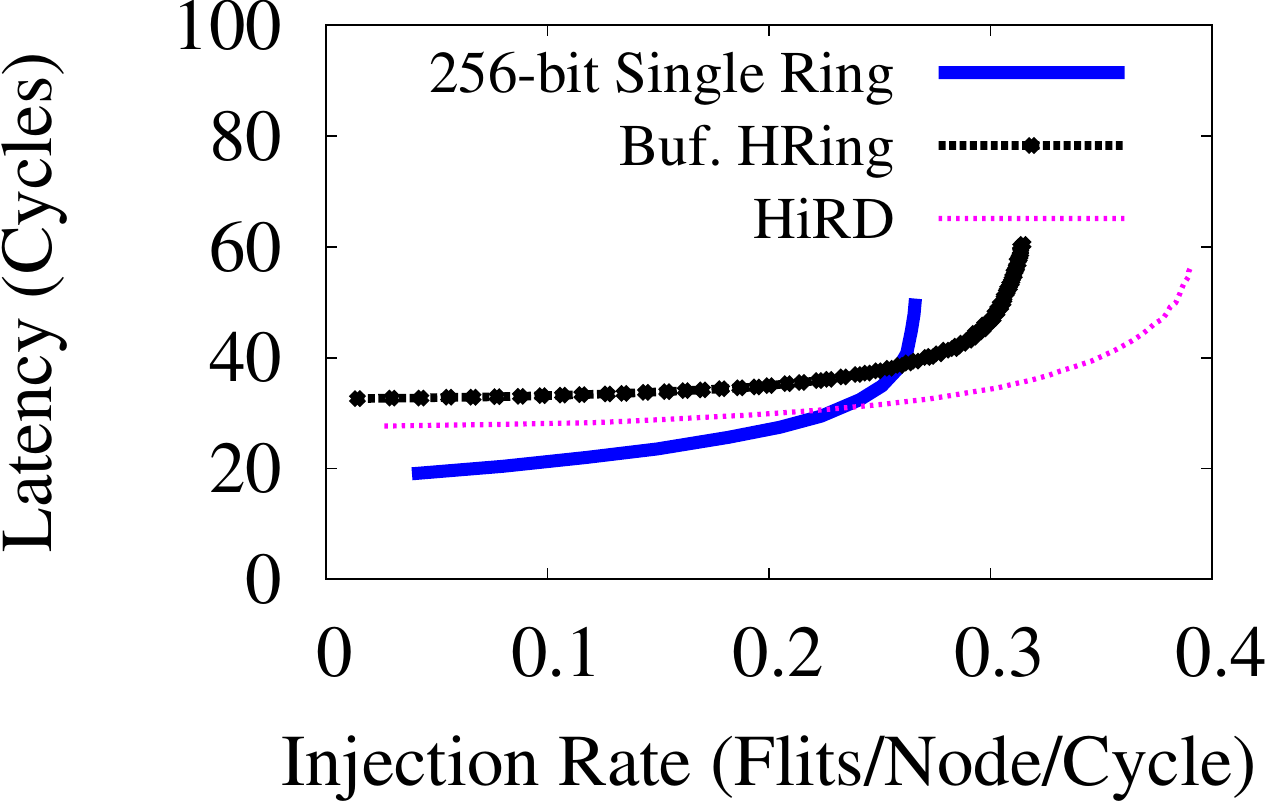}
\label{fig:synth8UR}
}
\\
\vspace{-0.1in}
\subfloat[Bit Complement, 4x4]{
\includegraphics[width=1.5in]{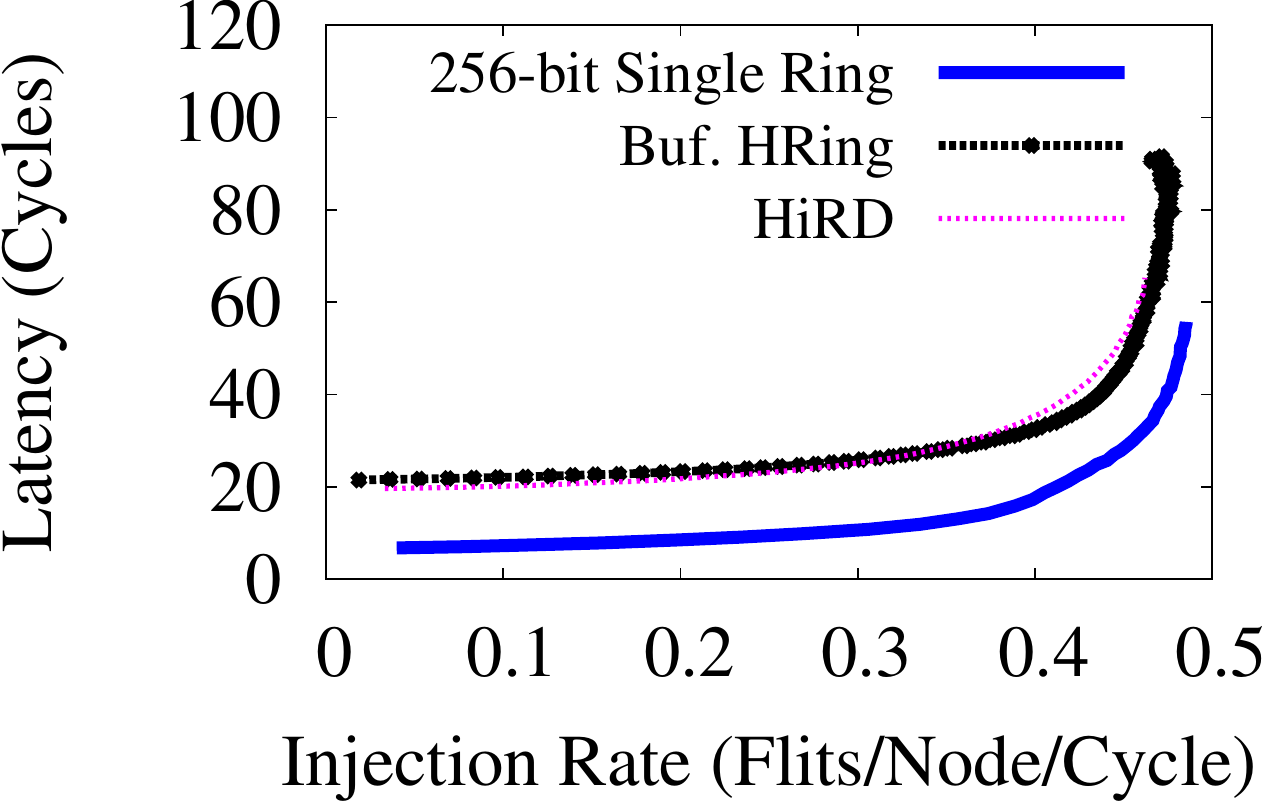}
\label{fig:synthBC}
}
\subfloat[Bit Complement, 8x8]{
\includegraphics[width=1.5in]{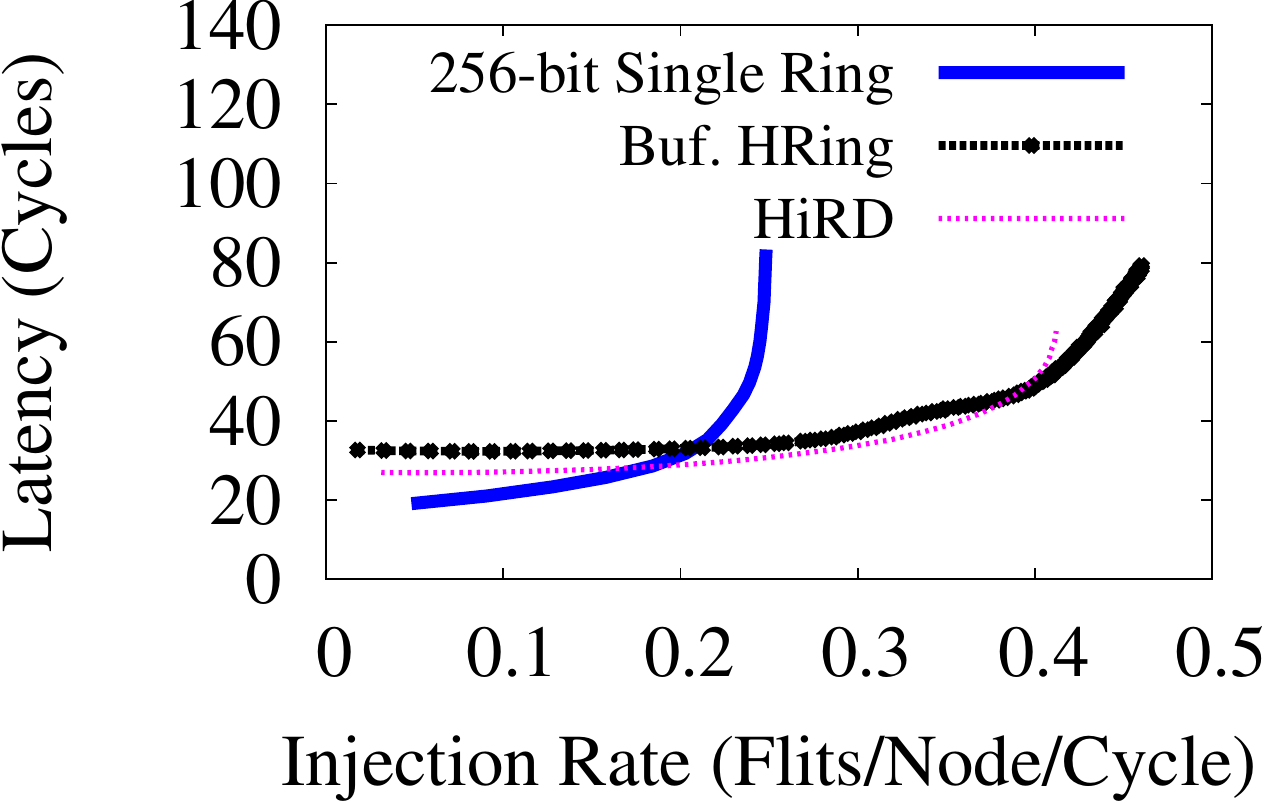}
\label{fig:synth8BC}
}
\\
\vspace{-0.1in}
\subfloat[Transpose, 4x4]{
\includegraphics[width=1.5in]{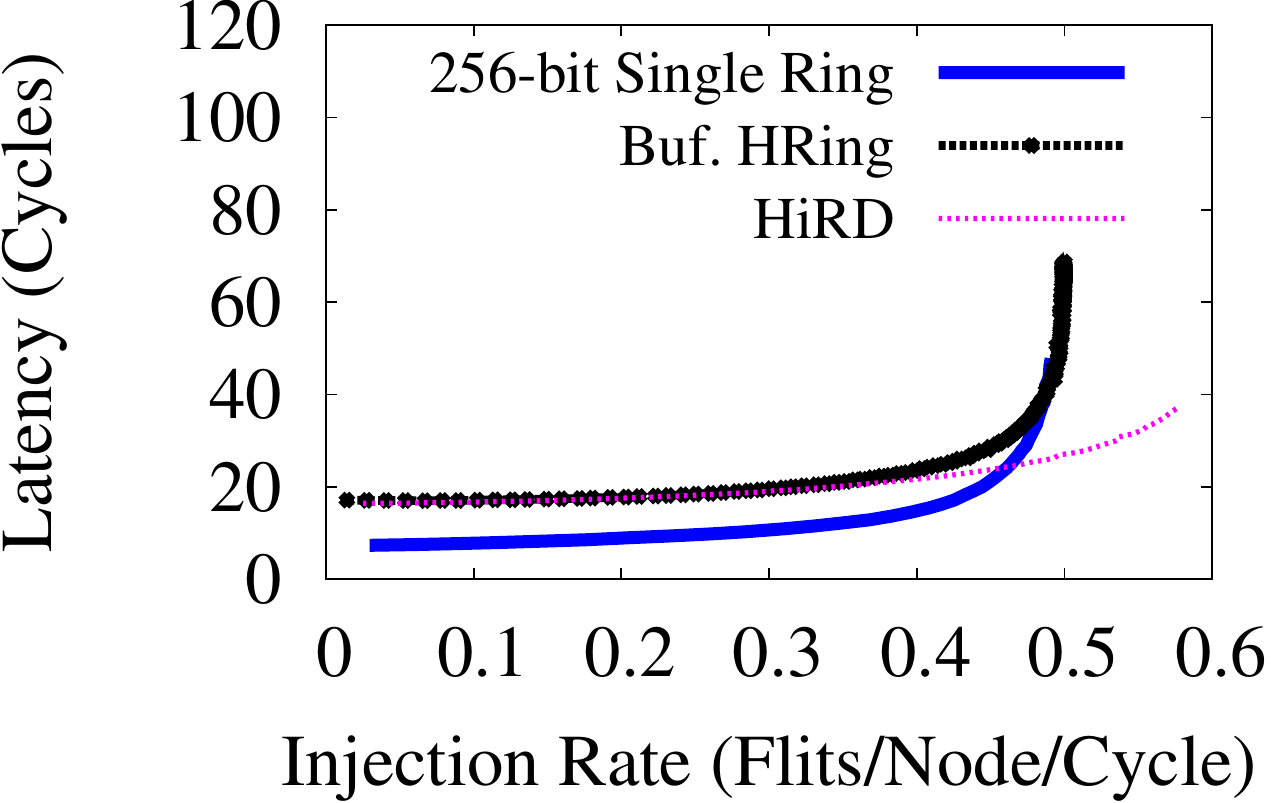}
\label{fig:synthTR}
}
\subfloat[Transpose, 8x8]{
\includegraphics[width=1.5in]{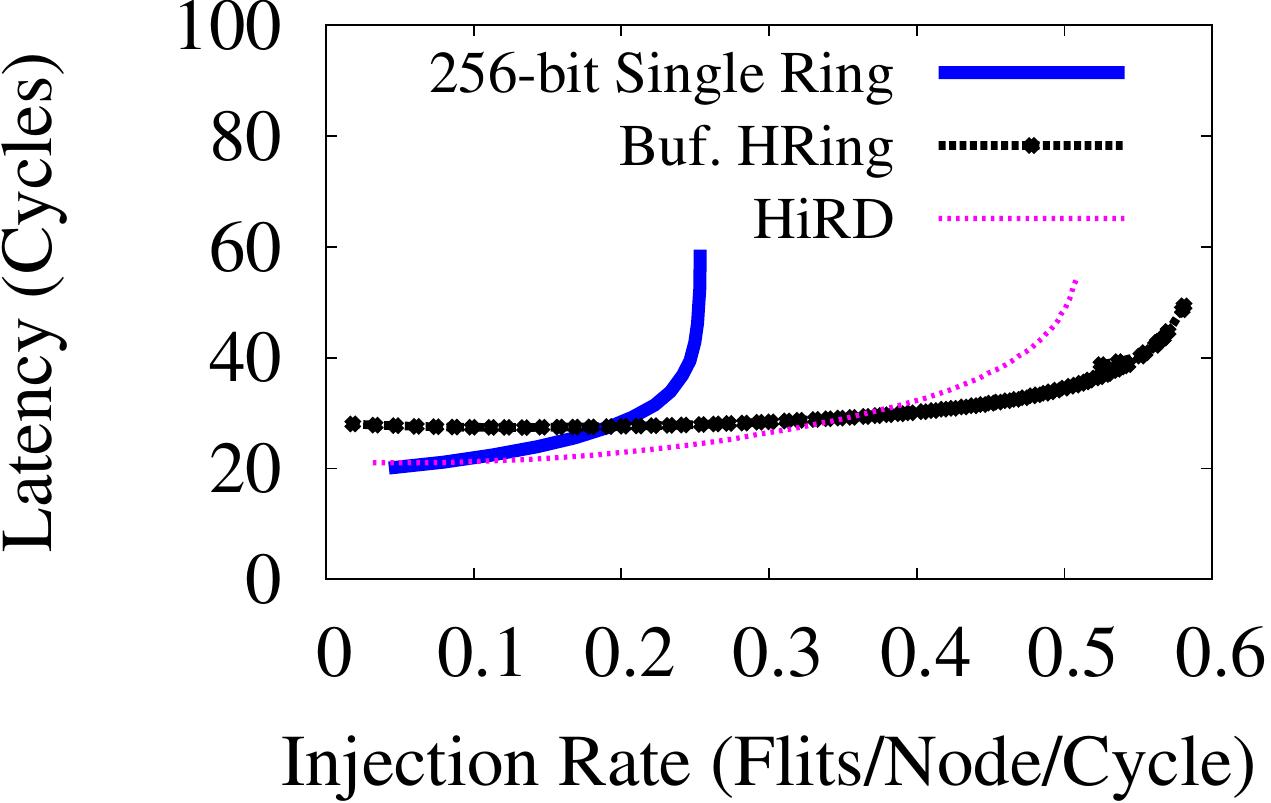}
\label{fig:synth8TR}
}
\vspace{-0.05in}
\caption{\footnotesize \bf Synthetic-traffic evaluations for 4x4 and 8x8 networks.}
\label{fig:synth}
\vspace{-0.15in}
\end{figure}

Figure~\ref{fig:synth} shows the average packet latency as a function of injection rate for
buffered and bufferless mesh routers, a single-ring design, the buffered
hierarchical ring, and HiRD in 16 and 64-node systems. We show uniform random,
transpose and bit complement traffic patterns~\cite{principles}. Sweeps on injection rate
terminate at network saturation. The buffered hierarchical ring saturates at a
similar point to HiRD but maintains a slightly lower average latency because it
avoids transfer deflections. In contrast to these high-capacity designs, the
256-bit single ring saturates at a lower injection rate. 

As network size scales to 8x8, HiRD performs significantly better than the
256-bit single ring, because the hierarchy reduces the cross-chip latency while
preserving bisection bandwidth.  HiRD also performs better than Buffered HRing
because of two reasons. First, HiRD is able to allow higher peak utilization
(91\%) than Buffered HRing (71\%) on the global rings. We observed that when
flits have equal distance in a clock-wise and counter clock-wise direction,
Buffered HRing has to send flits to one direction in order to avoid deadlock
while deflections in HiRD allow flits to travel in both directions, leading to
better overall network utilization. Second, at high injection rates, the
transfer guarantee~\cite{hird} starts throttling the network, disallowing
future flits to be injected into the network until the existing flits arrive at
their destinations. This reduces congestion in the network and allows HiRD to
saturate at a higher injection rate than the buffered hierarchical ring design.

\ignore{
Figure~\ref{fig:synth} shows average latency as a function of injection rate for
buffered and bufferless mesh routers, a single-ring design, the buffered
hierarchical ring, and HiRD in 16 and 64-node systems. We show uniform random,
transpose, and bit-complement traffic patterns. Sweeps terminate at network
saturation. For comparison, we also add a buffered mesh with buffer bypassing
and CHIPPER~\cite{chipper}, a bufferless mesh network. HiRD performs very
closely to the buffered mesh design in 4x4, saturating at nearly the same
injection rate but maintaining a slightly lower network latency at high load.
The buffered hierarchical ring saturates at a similar point to HiRD and the
buffered mesh, but maintains a slightly lower average latency because it avoids
transfer deflections. In contrast to these high-capacity designs, the single
ring saturates at a lower injection rate, closer to the bufferless mesh design,
CHIPPER. As network size scales to 8x8, HiRD performs significantly better
relative to the meshes, because the hierarchy reduces the cross-chip latency
while preserving bisection bandwidth.  Lastly, HiRD also performs better than
Buffered HRing because of two reasons. First, HiRD is able to allow higher
peak utilization (91\%) than Buffered HRing (71\%) on the global rings. We
observed that when flits have equal distance in a clock-wise and counter
clock-wise direction, Buffered HRing has to send flits to one direction in
order to avoid deadlock while deflections in HiRD allow flits to travel in
both directions, leading to better overall network utilization. Second, at high injection rates, the transfer guarantee~\cite{hird} starts
throttling the network, disallowing future flits to be injected into the network
until the existing flits arrive at their destinations. This reduces congestion
in the network and allow HiRD to saturate after a buffered 
hierarchical ring design.
}

\ignore{
\begin{figure}[h!]
\centering
\includegraphics[width=4in]{plots/synthHRStage.pdf}
\caption{\small Uniform Random synthetic traffic sweeping the number 
of pipeline stages in a hierarchical ring 8-bridge design}
\label{fig:synthStage}
\end{figure}
}

\subsection{Injection and Transfer Guarantees}
\label{sec:eval-guarantees}

In this subsection, we study HiRD's behavior under a worst-case
synthetic traffic pattern that triggers the injection and transfer
guarantees and demonstrates that they are necessary for correct
operation, and that they work as designed.

\noindent\textbf{Traffic Pattern.} In the worst-case traffic pattern, all nodes
on three rings in a two-level (16-node) hierarchy inject traffic (we call these
rings Ring A, Ring B, and Ring C). Rings A, B, and C have bridge routers
adjacent to each other, in that order, on the single global ring. All nodes in
Ring A continuously inject flits 
to nodes in Ring C, and all
nodes in Ring C likewise inject flits to nodes in Ring A. This creates heavy
traffic on the global ring across the point at which Ring B's bridge router
connects. All nodes on Ring B continuously inject flits (whenever they are
able) addressed to another ring elsewhere in the network. However, because
Rings A and C continuously inject flits, Ring B's bridge router will not be
able to transfer any flits to the global ring in the steady state (unless
another mechanism such as the throttling mechanism in~\cite{hird} intercedes).

\noindent\textbf{Results.} Table~\ref{table:guarantee-results} shows
three pertinent metrics on the network running the described traffic
pattern: average network throughput (flits/node/cycle) for nodes on
Rings A, B, and C, the maximum time (in cycles) spent by any one flit
at the head of a transfer FIFO, and the maximum number of times any
flit is deflected and has to circle a ring to try again. These metrics
are reported with the injection and transfer guarantee mechanisms
disabled and enabled. The experiment is run with the synthetic traffic
pattern for 300K cycles.

\linespread{0.95}
\begin{table*}
\vspace{-0.25in}
\small
\centering
\begin{tabular}{|l||l|l|l||l|l|}
\hline
Configuration &  \multicolumn{3}{|c||}{\breakTable{Network Throughput\\(flits/node/cycle)}} & \breakTable{Transfer FIFO Wait\\(cycles)} & Deflections/Retries \\
\hline
 & Ring A & Ring B & Ring C & avg/max & avg/max \\
\hline
\hline
Without Guarantees & 0.164 & 0.000 & 0.163 & 2.5 / 299670 & 6.0 / 49983 \\
\hline
With Guarantees & 0.133 & 0.084 & 0.121 & 1.2 / 66 & 2.8 / 18 \\
\hline
\end{tabular}
\caption{\footnotesize \bf Results of worst-case traffic pattern without and with
  injection/transfer guarantees enabled.}
\label{table:guarantee-results}
\end{table*}
\linespread{0.85}

The results show that \emph{without} the injection and transfer guarantees,
Ring B is completely starved and cannot transfer any flits onto the
global ring. This is confirmed by the maximum transfer FIFO wait time,
which is almost the entire length of the simulation. In other words,
once steady state is reached, no flit ever transfers out of Ring
B. Once the transfer FIFO in Ring B's bridge router fills, the local
ring fills with more flits awaiting a transfer, and these flits are
continuously deflected. Hence, the maximum deflection count is very
high.  Without the injection or transfer guarantees, the network does
\emph{not} ensure forward progress for these flits. In contrast, when
the injection and transfer guarantees are enabled, (i) Ring B's bridge
router is able to inject flits into the global ring and (ii) Ring B's
bridge router fairly picks flits from its local ring to place into its
transfer FIFO. The maximum transfer FIFO wait time and maximum
deflection count are now bounded, and nodes on all rings receive
network throughput.  Thus, the guarantees are both necessary
and sufficient to ensure deterministic forward progress for all flits
in the network. 

%


\noindent\textbf{Real Applications.} Table~\ref{table:guarantee-real-results}
shows the effect of the transfer guarantee mechanism on real applications in a
4x4 network.  Average transfer FIFO wait time shows the average number of cycles
that a flit waits in the transfer FIFO across all 60 workloads.
Maximum transfer FIFO wait time shows the maximum observed flit wait time in the same FIFO across all
workloads.  As illustrated in Table~\ref{table:guarantee-real-results}, some number of
flits can experience very high wait times when there is no transfer guarantee.
Our transfer guarantee mechanism reduces both average and maximum FIFO
wait times\footnote{As the network scales to 64 nodes, we observe that the average
wait time in the transfer FIFO does not affect the overall performance significantly
(adding 1.5 cycles per flit).}. In addition, we observe that our transfer
guarantee mechanism not only provides livelock- and deadlock-freedom but also
provides lower maximum wait time in the transfer FIFO for each flit because
the guarantee provides a form of throttling when the network is congested. A
similar observation has been made in many previous network-on-chip works
that use source throttling to improve the performance of the
network~\cite{selftuned,baydal05,hotnets2010,hat-sbac-pad,sigcomm12}.


\begin{table}[h]
\small
\centering
\begin{tabular}{|l||l|l|}
\hline
Configuration & Transfer FIFO Wait time (cycles) & Deflections/Retries \\
\hline
 & (avg/max) & (avg/max) \\
\hline
\hline
Without guarantees & 3.3 / 169 & 3.7 / 19 \\
\hline
With guarantees & 0.76 / 72 & 0.7 / 8 \\
\hline
\end{tabular}
\caption{\footnotesize \bf Effect of transfer guarantee mechanism on real workloads.}
\label{table:guarantee-real-results}
\end{table}


We conclude that our transfer guarantee mechanism is effective in eliminating
livelock and deadlock as well as reducing packet queuing delays in real workloads.

\subsection{Network Latency and Latency Distribution}



\begin{figure}[h!]
\centering
\includegraphics[width=5in]{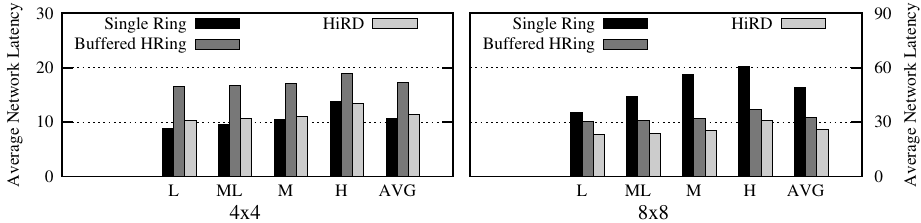}
\caption{\footnotesize \bf Average network latency for 4x4 and 8x8 networks.}
\label{fig:avg_Latency}
\end{figure}

Figure~\ref{fig:avg_Latency} shows average network latency for our three
evaluated configurations: 256-bit single ring, buffered hierarchical ring and
HiRD.  This plot shows that our proposal can reduce the network latency by
having a faster local-ring hop latency compared to other ring-based designs.
Additionally, we found that, for all real workloads, the number of deflections
we observed is always less than 3\% of the total number of flits. Therefore,
the benefit of our deflection based router design outweighs the extra cost of
deflections compared to other ring-based router designs. Finally, in the
case of small networks such as a 4x4 network, a 1-cycle hop latency of a single
ring provides significant latency reduction compared to the buffered
hierarchical design. However, a faster local-ring hop latency in HiRD helps to
reduce the network latency of a hierarchical design and provides a
competitive network latency compared to a single ring design in small
networks.


\begin{figure}[h!]
\centering
\vspace{-.1in}
\includegraphics[width=5in]{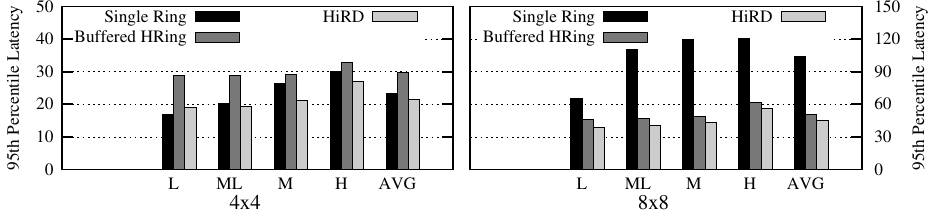}
\caption{\footnotesize \bf 95th percentile latency for 4x4 and 8x8 networks.}
\label{fig:95th_Latency}
\end{figure}

In addition, Figure~\ref{fig:max_Latency} shows the maximum latency 
and Figure~\ref{fig:95th_Latency} shows the 95th percentile
latency for each network design. The 95th percentile latency shows the behavior of the network without
extreme outliers. These two figures provide quantitative evidence
that the network is deadlock-free and livelock-free. Several
observations are in order:

\begin{figure}[h!]
\centering
\includegraphics[width=5in]{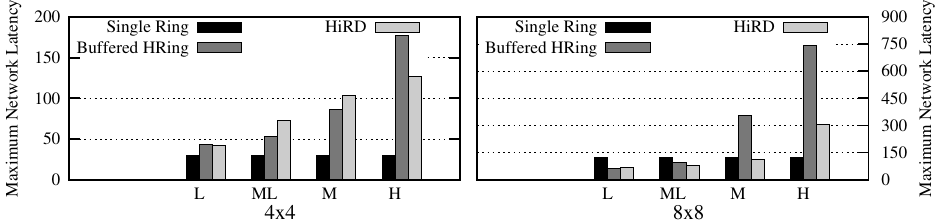}
\caption{\footnotesize \bf Maximum network latency for 4x4 and 8x8 networks.}
\vspace{-.1in}
\label{fig:max_Latency}
\end{figure}

1. HiRD provides lower latency at the 95th percentile and the lowest average
latency observed in the network. This lower latency comes from our transfer
guarantee mechanism, which is triggered when flits spend more than 100 cycles in
each local ring, draining all flits in the network to their destination. This
also means that HiRD improves the worst-case latency that a flit can
experience because none of the flits are severely delayed.

2. While both HiRD and the buffered hierarchical ring have higher 95th
percentile and maximum flit latency compared to a 64-bit single ring network,
both hierarchical designs have 60.1\% (buffered hierarchical ring) and 53.9\%
(HiRD) lower average network latency in an 8x8 network because a hierarchical design
provides better scalability on average. 

3. Maximum latency in the single ring is low because contention happens only at
injection and ejection, as opposed to hierarchical designs where contention can
also happen when flits travel through different level of the hierarchy.

4. The transfer guarantee in HiRD also helps to significantly reduce the maximum
latency observed by some flits compared to a buffered design because the
guarantee enables the throttling of the network, thereby alleviating congestion.
Reduced congestion leads to reduced maximum latency. This observation is confirmed by our synthetic
traffic results shown in Section~\ref{sec:eval-synth}.


\subsection{Fairness}

Figure~\ref{fig:hird-fairness} shows the fairness, measured by the maximum
slowdown metric, for our three evaluated configurations. Compared to a buffered
hierarchical ring design HiRD, is 8.3\% (5.1\%) more fair on a 4x4 (8x8)
network. Compared to a single ring design, HiRD is 40.0\% (296.4\%) more fair
on a 4x4 (8x8) network. In addition, we provide several observations:


\begin{figure*}[h!]
\vspace{-0.1in}
\centering
\includegraphics[width=5in]{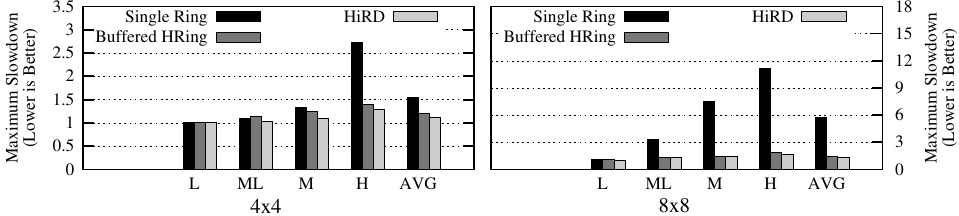}
\vspace{-0.05in}
\caption{\footnotesize \bf Unfairness for 4x4 and 8x8 networks.}
\label{fig:hird-fairness}
\end{figure*}


1. HiRD is the most fair design compared to the buffered hierarchical ring and
the single ring designs. Compared to a single ring design, hierarchical designs
are more fair because the global ring in the hierarchical designs allows flits
to arrive at the destination faster. Compared to the buffered hierarchical ring
design, HiRD is more fair because HiRD has lower average network latency. HiRD
is much more fair for medium and high intensity workloads, where the throttling
mechanism in HiRD lowers average network latency. 

2. Global rings allow both hierarchical designs to provide better fairness
compared to the single ring design as the size of the network gets bigger from
4x4 to 8x8. 

3. We conclude that HiRD is the most fair ring design among all evaluated
designs due to its overall lower packet latencies and reduced congestion across
all applications.

\subsection{Router Area and Timing}
\label{sec:eval-area}

We show both critical path length and normalized die area for single-ring,
buffered hierarchical ring, and HiRD, in Table~\ref{table:timing_area}. Area
results are normalized to the buffered hierarchical ring baseline, and are
reported for all routers required by a 16-node network (e.g., for HiRD, 16 node
routers and 8 bridge routers).

\linespread{1}
\begin{table}[h!]
\centering
\vspace{-0.1in}
\footnotesize{
\begin{tabular}{|p{1.9cm}||p{15mm}|p{20mm}|p{10mm}|}
\hline
Metric & Single-Ring & Buffered HRing & HiRD \\
\hline
\hline
Critical path (ns) & 0.33 & 0.87 & 0.61 \\
\hline
Normalized area & 0.281 & 1 & 0.497 \\
\hline
\end{tabular}
}
\vspace{-0.05in}
\caption{\footnotesize \bf Total router area (16-node network) and critical path.}
\label{table:timing_area}
\vspace{-0.15in}
\end{table}

\linespread{0.85}

Two observations are in order. First, HiRD reduces area relative to the
buffered hierarchical ring routers, because the node router required at each
network node is much simpler and does not require complex flow control logic.
HiRD reduces total router area by 50.3\% vs. the buffered hierarchical ring.
Its area is higher than a single ring router because it contains buffers in
bridge routers. However, the energy efficiency of HiRD and its performance at
high load make up for this shortcoming. Second, the buffered
hierarchical ring router's critical path is 42.6\% longer than HiRD because its control
logic must also handle flow control (it must check whether credits are
available for a downstream buffer).  The single-ring network has a higher
operating frequency than HiRD because it does not need to accommodate ring
transfers (but recall that this simplicity comes at the cost of poor
performance at high load for the single ring).

\ignore{
We show both critical path length and normalized die area for buffered and
bufferless mesh, single-ring, buffered hierarchical ring, and HiRD, in
Table~\ref{table:timing_area}. Area results are normalized by the buffered mesh
baseline, and are reported for all routers required by a 16-node network (e.g.,
for HiRD, 16 node routers and 8 bridge routers). 

Two observations are in order. First, hierarchical ring designs in
general significantly reduce area relative to the buffered hierarchical ring routers,
because the node router required at each network node is much simpler
and does not require complex flow control logic. HiRD
reduces total router area by 50.3\% vs. the buffered mesh. Its area
is higher than a single ring router because it contains
buffers in bridge routers. However, the energy efficiency of HiRD
and its performance at high load make up for this shortcoming. Second,
unlike HiRD's, the buffered hierarchical ring router's critical path is 42.6\%
longer  because its control logic must also handle flow control (it must check
whether credits are available for a downstream buffer).
The single-ring network has a higher operating frequency than HiRD
because it does not need to accommodate ring transfers (but recall
that this simplicity comes at the cost of poor performance at high
load for the single ring).
}

\ignore{
\begin{table}[h!]
\centering
\footnotesize{
\begin{tabular}{|p{1.5cm}||p{8mm}|p{10mm}|p{7mm}|p{8mm}|p{7mm}|}
\hline
Metric & Buffered & CHIPPER & Single-Ring & Buffered HRing & HiRD \\
\hline
\hline
Critical path length (ns) & 1.21 & 0.93 & 0.33 & -- & 0.61 \\
\hline
Normalized area & 1 & 0.181 & 0.205 & 0.729 & 0.362 \\
\hline
\end{tabular}
}
\caption{\footnotesize \bf Total router area (16-node network) and critical path.}
\label{table:timing_area}
\vspace{-0.3in}
\end{table}
}

\subsection{Sensitivity to Link Bandwidth}

\begin{figure}[h!]
\centering
\includegraphics[width=5in]{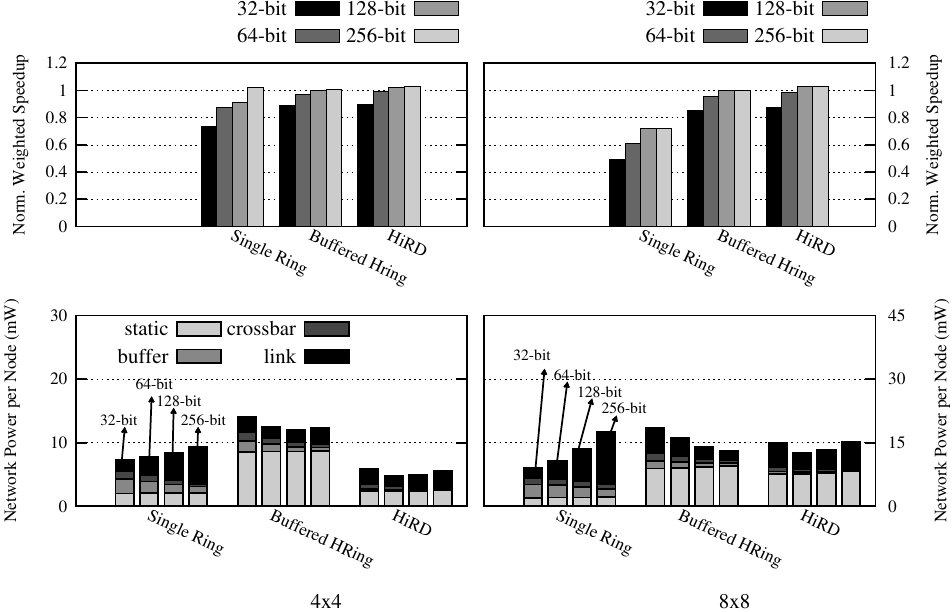}
\caption{\footnotesize \bf Sensitivity to different link bandwidth for 4x4 and 8x8 networks.}
\label{fig:bw_sweep}
\end{figure}

The bandwidth of each link also has an effect on the performance of different
network designs. We evaluate the effect of different link
bandwidths on several ring-based networks by using 32-, 64- and 128-bit links on all
network designs. Figure~\ref{fig:bw_sweep} shows the performance and power
consumption of each network design. As links get
wider, the performance of each design increases.  According to the evaluation results, HiRD performs
slightly better than a buffered hierarchical ring design for almost all link
bandwidths while maintaining much lower power consumption on a 4x4 network, and
slightly lower power consumption on an 8x8 network. 


Additionally, we observe that increasing link bandwidth can decrease the network
power in a hierarchical design because lower link bandwidth causes more
congestion in the network and leads to more dynamic buffer, crossbar and link power
consumption due to additional deflections at the buffers. 
As the link bandwidth increases, congestion reduces, lowering
dynamic power. However, we observe that past a certain link bandwidth (e.g., 128 bits for
buffered hierarchical ring and HiRD), congestion no longer reduces, because
deflections at the buffers become the bottleneck instead. This leads to
diminishing returns in performance yet increased dynamic power.
\ignore{We also observe that this behavior saturates once the link
bandwidth in large enough (128-bit for buffered hierarchical ring and HiRD),
leading to no further performance gain and increased dynamic power because
links that are too large do not reduce the overall deflections
at the buffer any further. 
In this scenario, we observe that more energy is
required to transfer a packet to the destination, but the time a packet takes 
to reach its destination (packet latency) decreases. As power is energy over time,
wide links lead to higher average power consumption.}

\subsection{Sensitivity to Configuration Parameters}
\label{sec:eval-sensitivity}

\begin{figure}[h]
\centering
\subfloat[\scriptsize Number of bridge routers]{
\includegraphics[width=1.25in]{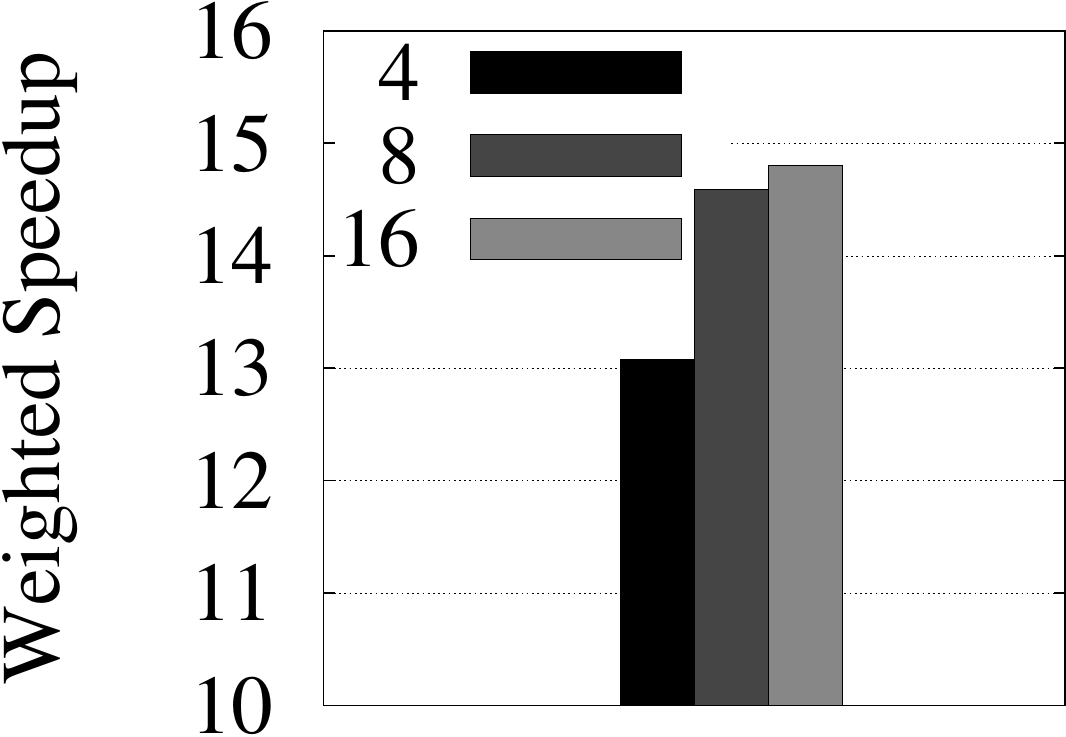}
\label{fig:bridges}
}
\subfloat[\scriptsize Local-to-global bridge buffer]{
\includegraphics[width=1.25in]{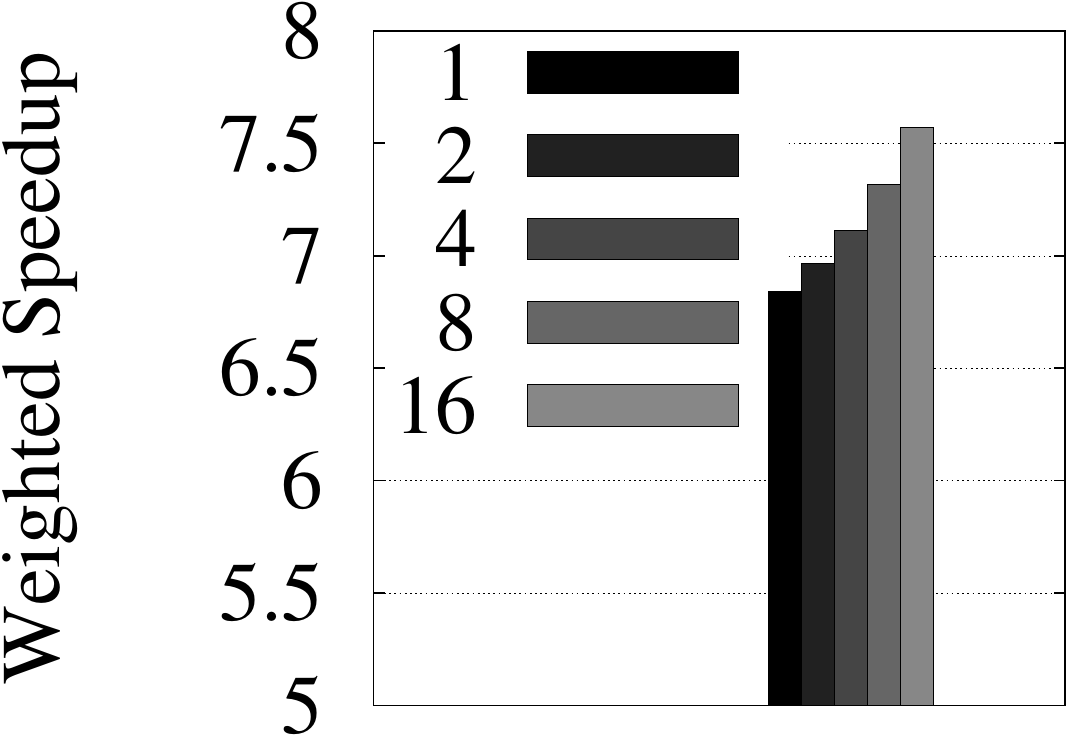}
\label{fig:L2GBuf}
}
\subfloat[\scriptsize Global-to-local bridge buffer]{
\includegraphics[width=1.25in]{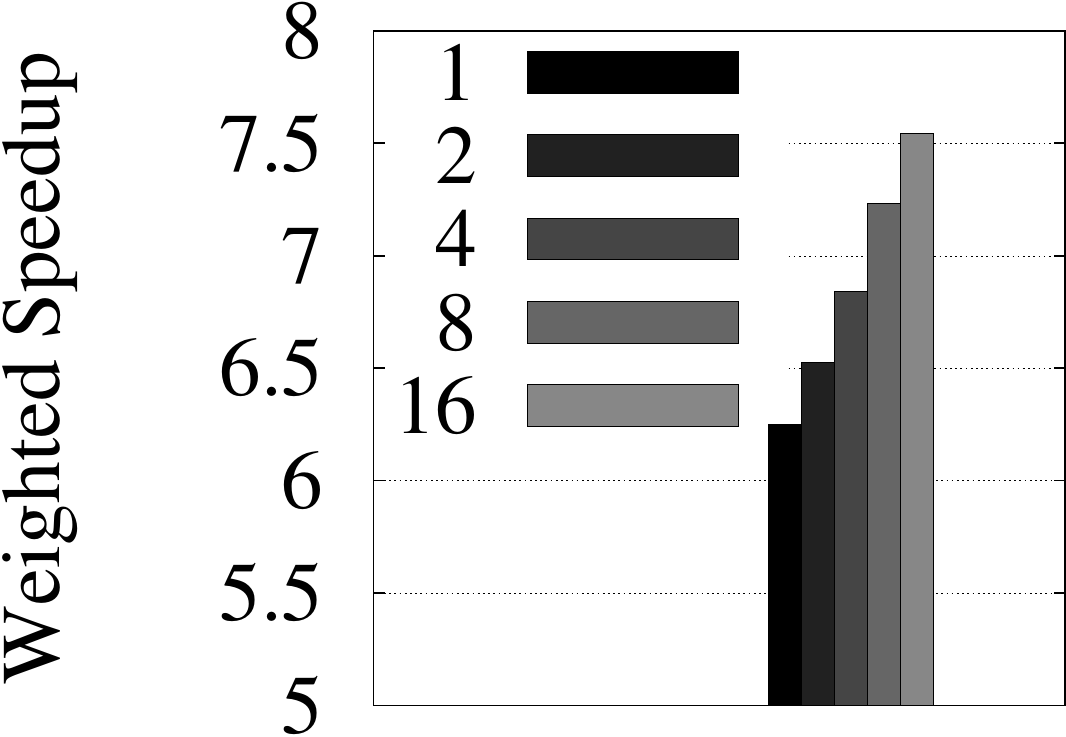}
\label{fig:G2LBuf}
}
\subfloat[\scriptsize Global-Local ring B/W ratio]{
\includegraphics[width=1.25in]{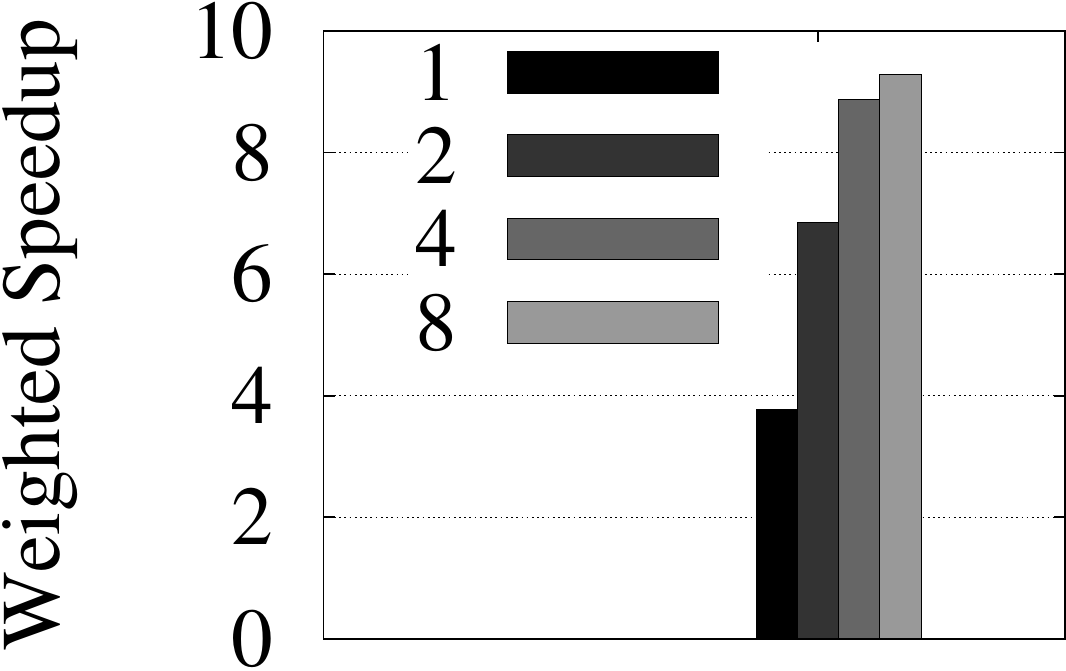}
\label{fig:GWidth}
}
\caption{\footnotesize \bf Performance sensitivity to buffer sizes and the global ring
  bandwidth in a 4x4 network.}
\end{figure}

\noindent\textbf{Bridge Router Organization.} The number of bridge
routers connecting the global ring(s) to the local rings has an
important effect on system performance because the connection between
local and global rings can limit bisection bandwidth. In
Figure~\ref{fig:topology}, we showed three alternative arrangements for
a 16-node network, with 4, 8, and 16 bridge routers. So far, we have
assumed an 8-bridge design in 4x4-node systems, and a system with 8
bridge routers at each level in 8x8-node networks
(Figure~\ref{fig:scale}). In Figure~\ref{fig:bridges}, we show average
performance across all workloads for a 4x4-node system with 4, 8, and
16 bridge routers. Buffer capacity is held constant. As shown,
significant performance is lost if only 4 bridge routers are used
(10.4\% on average). However, doubling from 8 to 16 bridge routers
gains only 1.4\% performance on average. Thus, the 8-bridge design
provides the best tradeoff of performance and network cost (power and
area) overall in our evaluations.


\noindent\textbf{Bridge Router Buffer Size.} The size of the FIFO queues used
to transfer flits between local and global rings can have an impact on
performance if they are too small (and hence are often full, leading them to
deflect transferring flits) or too large (and hence increase bridge router
power and die area). We show the effect of local-to-global and global-to-local
FIFO sizes in Figures~\ref{fig:L2GBuf}~and~\ref{fig:G2LBuf}, respectively, for
the 8-bridge 4x4-node design. In both cases, increased buffer size leads to
increased performance. However, performance is more sensitive to
global-to-local buffer size (20.7\% gain from 1-flit to 16-flit buffer size)
than to local-to-global size (10.7\% performance gain from 1 to 16 flits),
because in the 8-bridge configuration, the whole-loop latency around the global
ring is slightly higher than the loop latency in each of the local ring, making
a global-to-local transfer retry more expensive than a local-to-global one.


For our evaluations, we use a 4-flit global-to-local and 1-flit
local-to-global buffer per bridge router, which results in transfer deflection
rates of 28.2\% (global-to-local) and 34\% (local-to-global) on average for
multiprogrammed workloads. \green{These deflection rates are less than 1\% for
all of our multithreaded workloads. The deflection rate is much lower in multithreaded
workloads because these workloads are less memory-intensive and hence the contention
in the on-chip interconnect is low for them.}


\noindent\textbf{Global Ring Bandwidth.} Previous work on
hierarchical ring designs does not examine the impact of global ring
bandwidth on performance but instead assume equal bandwidth in
local and global rings~\cite{ravindran98}. In Figure~\ref{fig:GWidth},
we examine the sensitivity of system performance to global ring
bandwidth relative to local ring bandwidth, for the all-High category
of workloads (in order to stress check bisection bandwidth). Each point in
the plot is described by this global-to-local ring bandwidth ratio. The local ring design
is held constant while the width of the global ring is adjusted. If a
ratio of 1:1 is assumed (leftmost bar), performance is significantly
worse than the best possible design. Our main evaluations in 4x4
networks use a ratio of 2:1 (global:local) in order to provide
equivalent bisection bandwidth to a 4x4 mesh baseline. Performance
increases by 81.3\% from a 1:1 ratio to the 2:1 ratio that we
use. After a certain point, the global ring becomes less of a
bottleneck, and further global-ring bandwidth increases have massively smaller
effects. \ignore{Nevertheless, a high-ratio design point might be
  desirable to ameliorate concerns of worst-case performance under
  very high network load, at the cost of additional global ring area
  and energy.}

\noindent\textbf{Delivery Guarantee Parameters.}
We introduced injection guarantee and ejection guarantee mechanisms to ensure
every flit is eventually delivered to its destination. These guarantees are clearly described in detail in our original work~\cite{hird}. The
injection guarantee mechanism takes a threshold parameter that specifies how
long an injection can be blocked before action is taken.  Setting this
parameter too low can have an adverse impact on performance, because the system
throttles nodes too aggressively and thus underutilizes the network. Our main
evaluations use a $100$-cycle threshold. For high-intensity workloads,
performance drops by 21.3\% when using an aggressive threshold of only $1$ cycle. From $10$
cycles upward, variation in performance is at most 0.6\%: the mechanism is
invoked rarely enough that the exact threshold does not matter, only that it is
finite (is required for correctness guarantees). In fact, for a 100-cycle threshold, the injection
guarantee mechanism is \emph{never} triggered in our real applications. Hence, the
mechanism is necessary only for corner-case correctness. In addition, we
evaluate the impact of communication latency between routers and the
coordinator. We find less than 0.1\% variation in performance for latencies
ranging from $1$ to $30$ cycles (when parameters are set so that the mechanism becomes
active); thus, slow, low-cost wires may be used for this mechanism.

The ejection guarantee takes a single threshold parameter: the number
of times a flit is allowed to circle around a ring before action is
taken. We find less than 0.4\% variation in performance when sweeping the
threshold from $1$ to $16$.  Thus, the mechanism provides correctness
in corner cases but is unimportant for performance in the common case.


\subsection{Comparison Against Other Ring Configurations}
\label{sec:eval-ws-power}

\begin{figure}[h!]
\vspace{-.35in}
\centering
\hbox{
\includegraphics[width=5in]{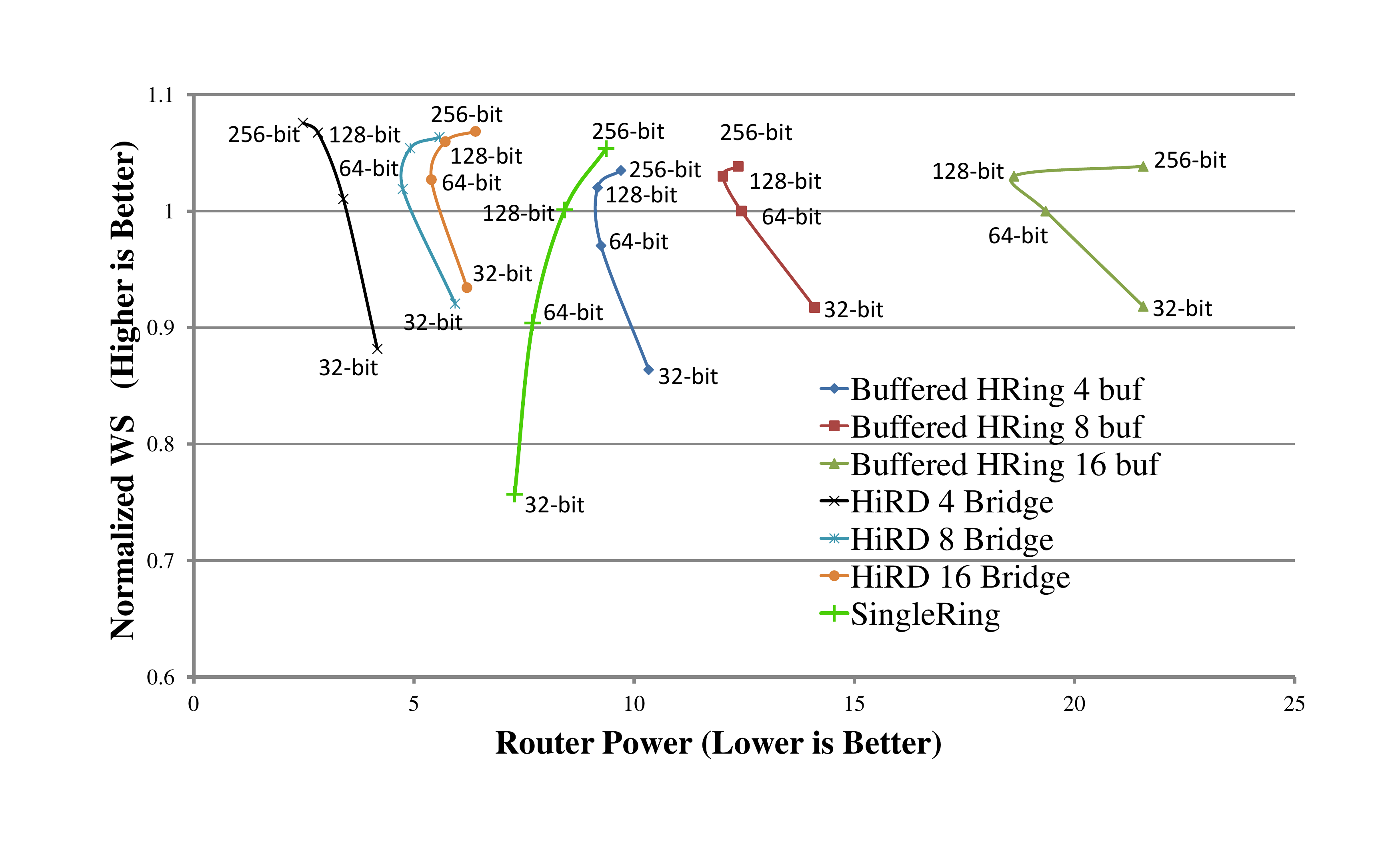}
}
\vspace{-.3in}
\caption{\footnotesize \bf Weighted speedup (Y) vs. power (X) for 4x4 networks.}
\vspace{-.15in}
\label{fig:WS_power}
\end{figure}

Figure~\ref{fig:WS_power} highlights the energy-efficiency comparison of
different ring-based design configurations by showing weighted speedup (Y axis)
against power (X axis) for all evaluated 4x4 networks. HiRD is shown with the
three different bridge-router configurations (described in
\S\ref{sec:bridge_router}). Every ring design is evaluated at various
link bandwidths (32-, 64-, 128- and 256-bit links). The top-left is the ideal
corner (high performance, low power). As the results show, at the same link
bandwidth, all three configurations of HiRD are more energy efficient than the
evaluated buffered hierarchical ring baseline designs at this network size.

We also observe that increasing link bandwidth can sometimes decrease router
power as it reduces deflections in HiRD or lowers contention at the buffers in
a buffered hierarchical ring design. However, once links are wide enough, this
benefit diminishes for two reasons: 1) 
links and crossbars consume more energy, 2)
packets arrive at the destination faster, leading to higher power as more energy
is consumed in less time.


\ignore{Additionally, we provide two additional comparison points against two flattened
butterfly designs. First, compared to the flatten butterfly design that has
equal bisection bandwidth (FbFly Eq. BW), HiRD performs 8.14\% better and also
consumes 53\% less power. Compared to the 64-bit link flatten butterfly, HiRD
performs 2.69\% slower, but consumes 58\% less power. }

\subsection{Comparison Against Other Network Designs}
\label{sec:eval-topos}

For completeness, Table~\ref{table:results-topo} compares HiRD against several other network
designs on 4x4 and 8x8 networks using the multiprogrammed workloads described in
Section~\ref{sec:meth}.


\ignore{
For completeness, Figure~\ref{fig:synth-topo} shows the comparison against several other network
designs on a 4x4 and 8x8 network using a synthetic workloads similar to
Section~\ref{sec:eval-synth}.

\begin{figure}[h!]
\centering
\subfloat[Uniform Random, 4x4]{
\includegraphics[width=1.75in]{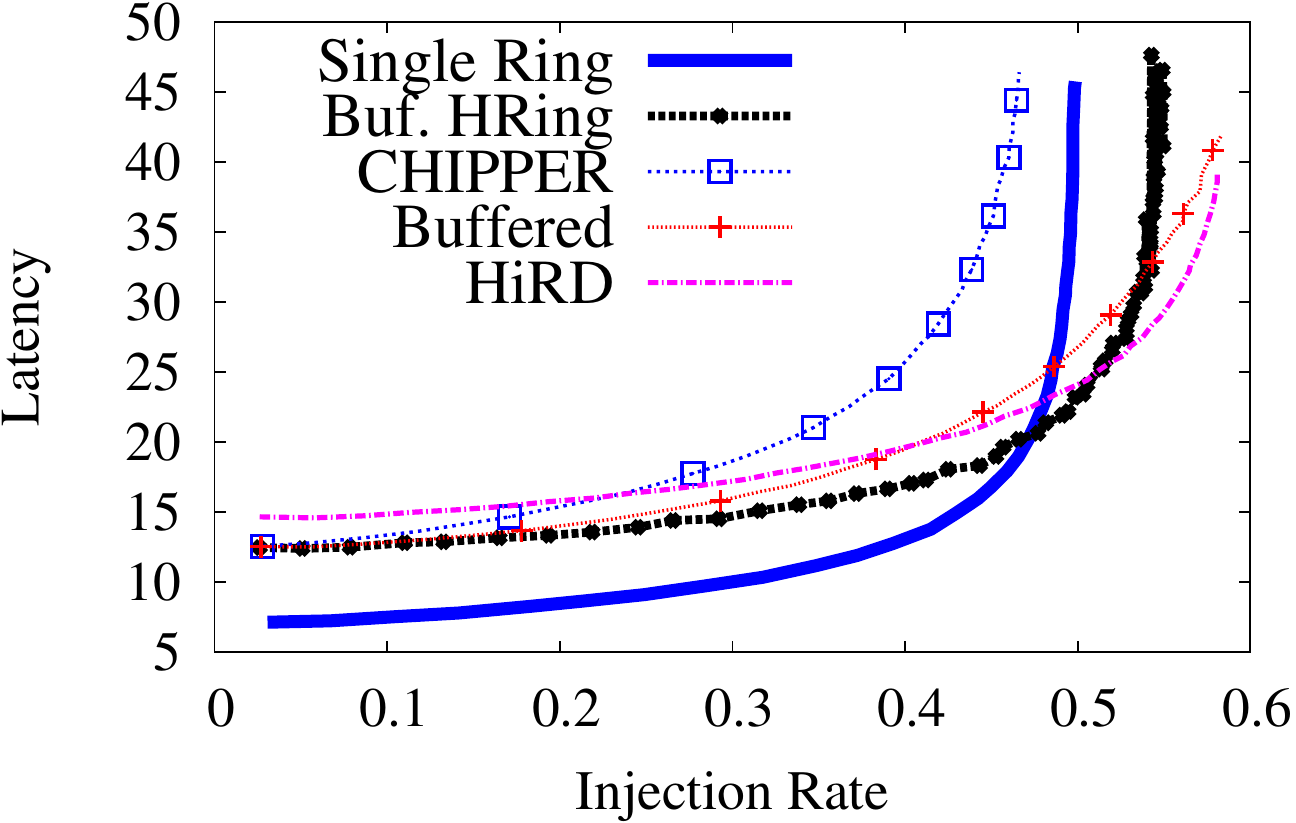}
\label{fig:synthUR-topo}
}
\subfloat[Bit Complement, 4x4]{
\includegraphics[width=1.75in]{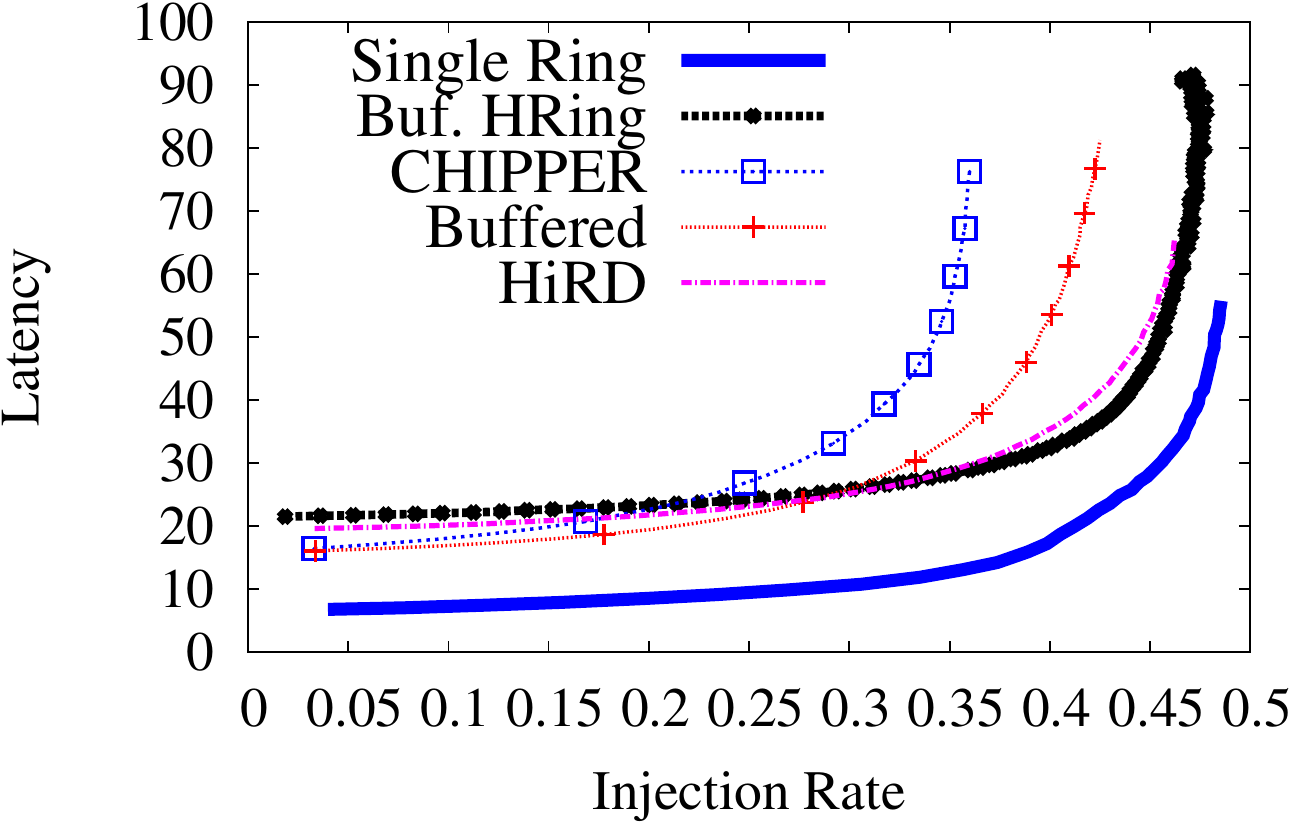}
\label{fig:synthBC-topo}
}
\subfloat[Transpose, 4x4]{
\includegraphics[width=1.75in]{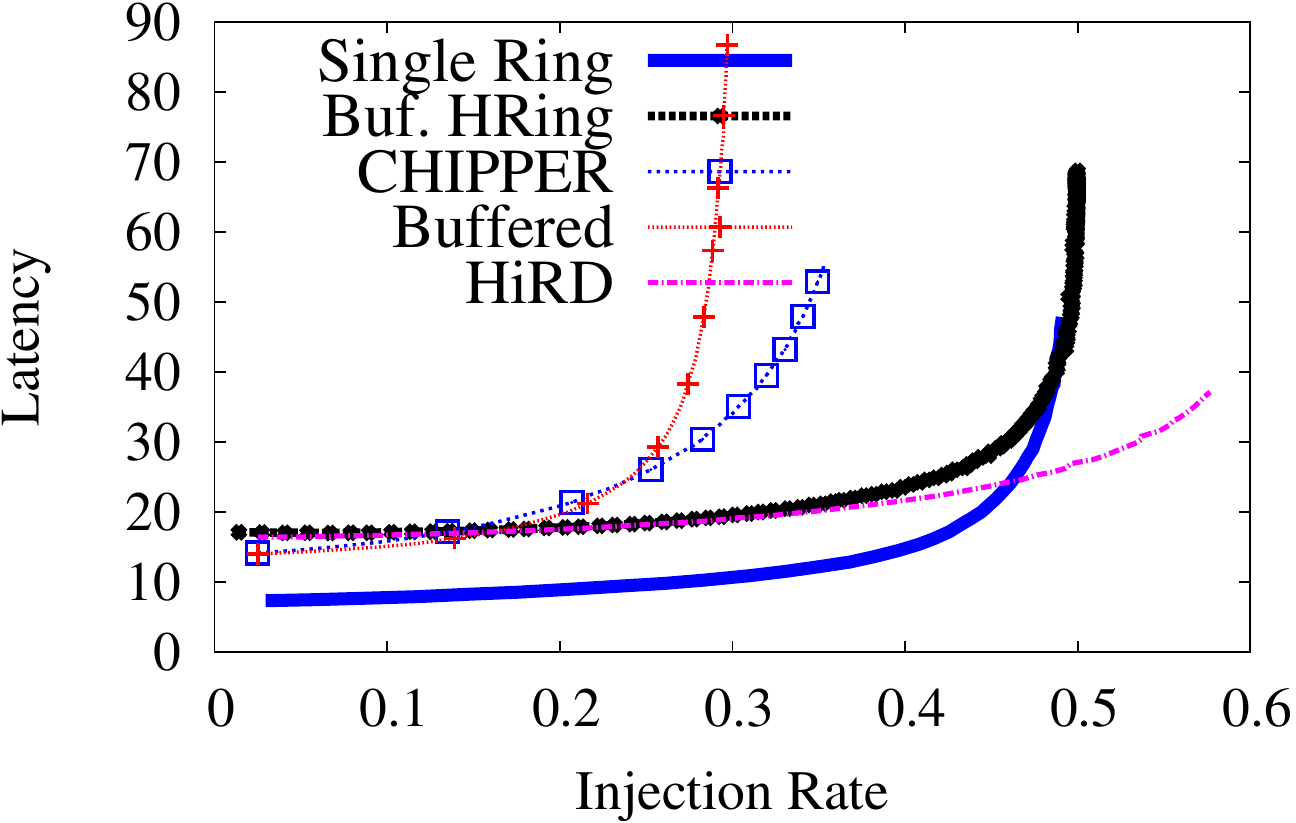}
\label{fig:synthTR-topo}
}
\\
\subfloat[Uniform Random, 8x8]{
\includegraphics[width=1.75in]{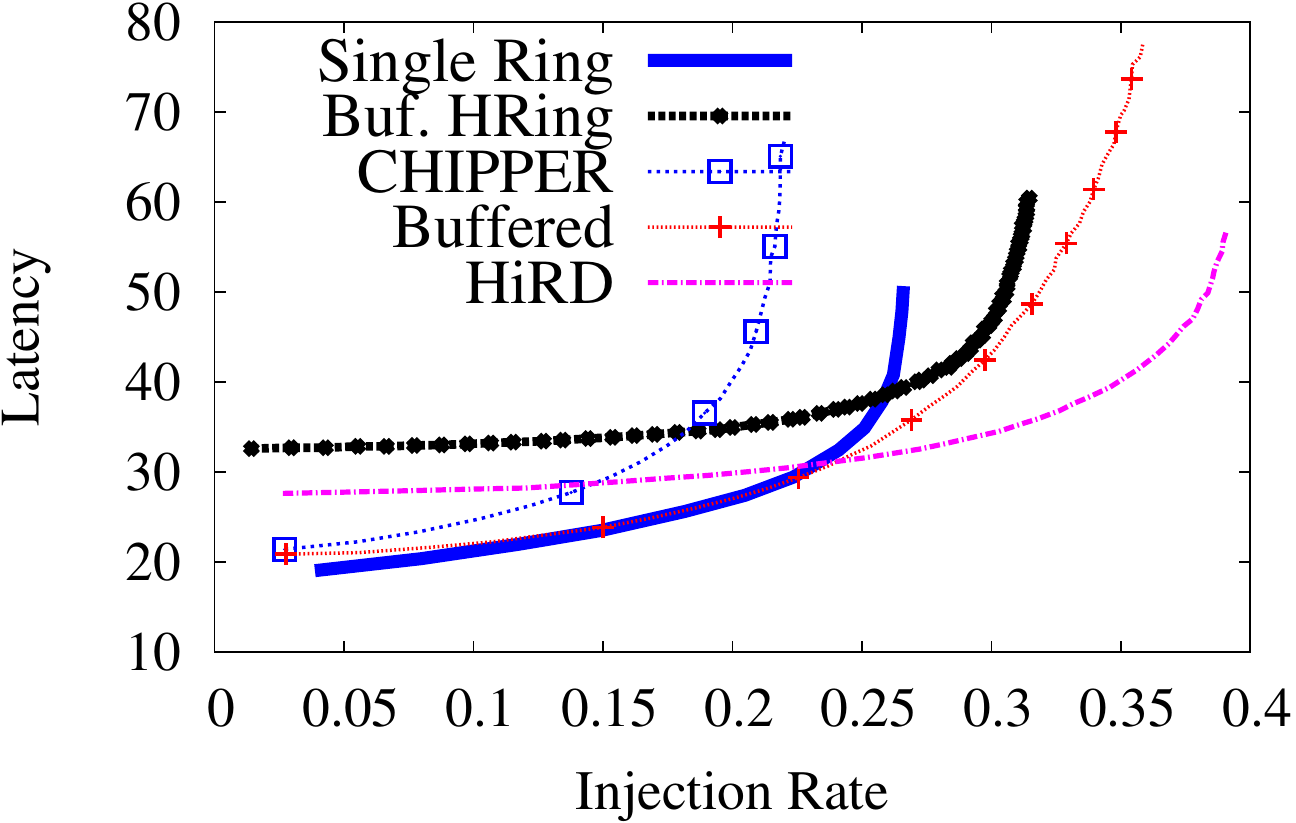}
\label{fig:synth8UR-topo}
}
\subfloat[Bit Complement, 8x8]{
\includegraphics[width=1.75in]{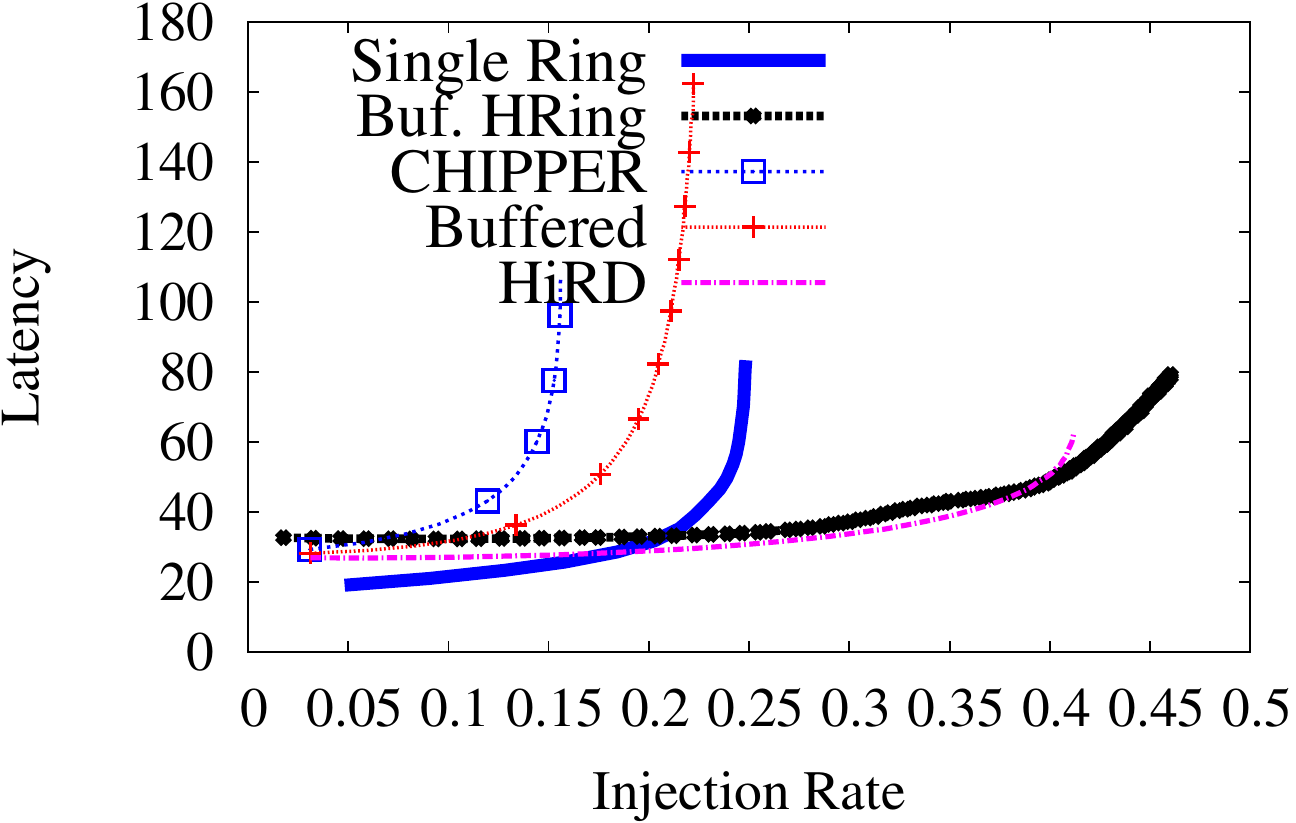}
\label{fig:synth8BC-topo}
}
\subfloat[Transpose, 8x8]{
\includegraphics[width=1.75in]{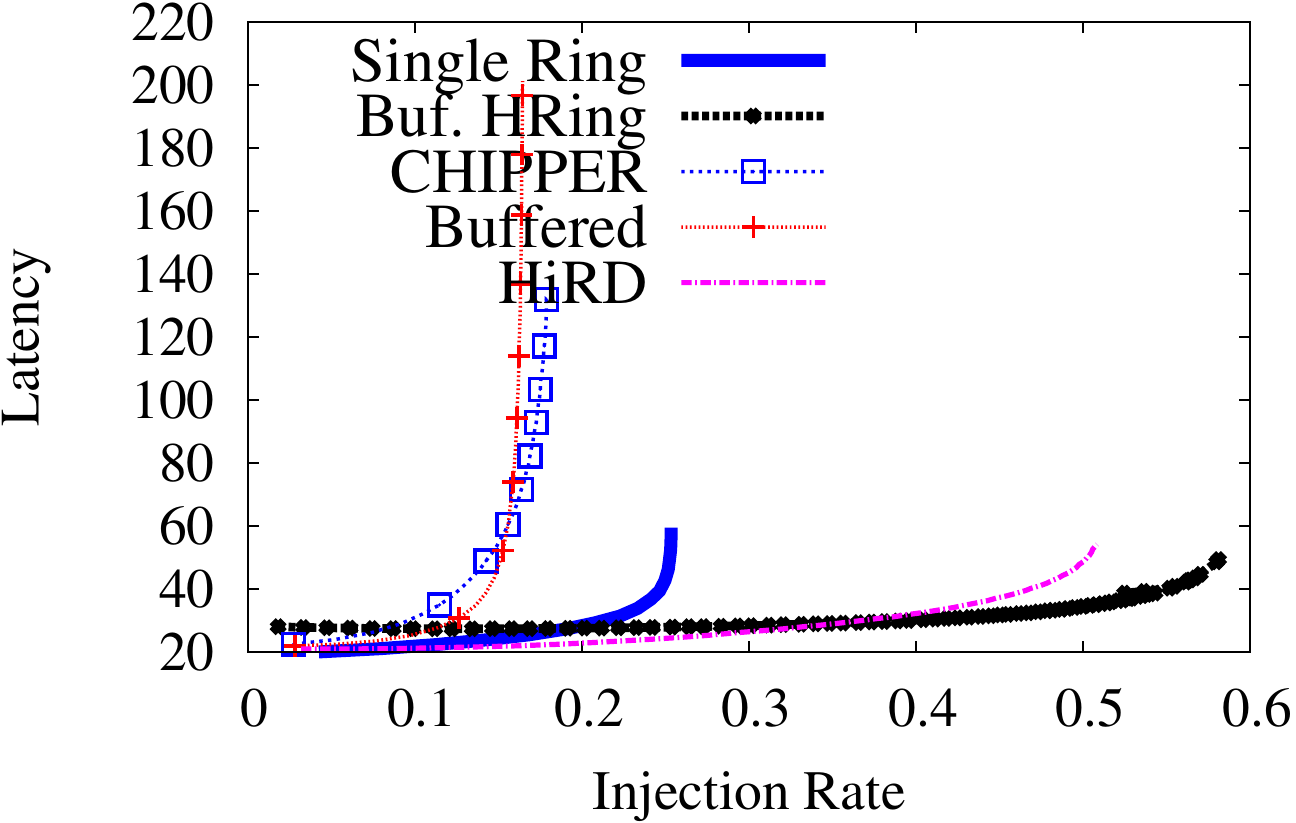}
\label{fig:synth8TR-topo}
}
\caption{\small Synthetic-traffic evaluations for 4x4 and 8x8 networks against different topologies.}
\label{fig:synth-topo}
\end{figure}
}

We compare our mechanism against a buffered mesh design with buffer bypassing~\cite{wang03,michelog10}.
We configure the buffered mesh to have 4 virtual channels (VCs) per port with 8
buffers per VC.  We also compare our mechanism against CHIPPER~\cite{chipper},
a low-complexity bufferless mesh network. We use 128-bit links for both designs.
Additionally, we compare our mechanism against a flattened
butterfly~\cite{flattened_bfly} with 4 VCs per output port, 8 buffers per VC,
and 64-bit links. Our main conclusions are as follows:


\begin{table}[h!]
\small
\centering
\begin{tabular}{|l||l|l||l|l|}
\hline
Topologies & \multicolumn{2}{c||}{4x4} & \multicolumn{2}{|c|}{8x8} \\
\hline
 & Norm. WS & Power (mWatts) & Norm. WS & Power (mWatts) \\
\hline
\hline
Single Ring & 0.904 & 7.696 & 0.782 & 13.603 \\
\hline
Buffered HRing & 1 & 12.433 & 1 & 16.188 \\
\hline
Buffered Mesh & 1.025 & 11.947 & 1.091 & 13.454 \\
\hline
CHIPPER & 0.986 & 4.631 & 1.013 & 7.275 \\
\hline
Flattened Butterfly & 1.037 & 10.760 & 1.211 & 30.434 \\
\hline
HiRD & 1.020 & 4.746 & 1.066 & 12.480 \\
\hline
\end{tabular}
\caption{\footnotesize \bf Evaluation for 4x4 and 8x8 networks against different network designs.}
\label{table:results-topo}
\end{table}



\indent 1. Against designs using the mesh topology, we observe that HiRD performs very
closely to the buffered mesh design both for 4x4 and 8x8 network sizes, while a
buffered hierarchical ring design performs slightly worse compared to HiRD and
buffered mesh designs. Additionally, HiRD performs better than CHIPPER in both
4x4 and 8x8 networks, though CHIPPER consumes less power in an 8x8 design as there
is no buffer in CHIPPER. \ignore{Rather, the single ring saturates at a lower injection
rate, closer to the bufferless mesh design, CHIPPER. As network size scales to 8x8,
HiRD performs significantly better relative to the meshes, because the
hierarchy reduces the cross-chip latency while preserving bisection bandwidth.}

2. Compared to a flattened butterfly design, we observe that HiRD performs
competitively with a flattened butterfly in a 4x4 network, but consumes lower
router power. In an 8x8 network, HiRD does not scale as well as a flattened
butterfly network and performs 11\% worse than a flattened butterfly network;
however, HiRD consumes 59\% less power than the flattened butterfly design. 

3. Overall, we conclude that HiRD is competitive in performance with the
highest performing designs while having much lower power consumption.


\ignore{
We compared our mechanism against a buffered mesh design with buffer bypassing.
We configure the buffered mesh to have 4 VCs per port with 8 buffers per VC. 
We also compare our mechanism against CHIPPER~\cite{chipper}, 
a bufferless mesh network. We use 128-bit links on both design. 
\ignore{Additionally,
we compared our mechanism against a flattened butterfly design topology~\cite{flattened_bfly} with
4 VCs per output port, 8 buffered per VC and 64-bit links.} We observe that HiRD performs very
closely to the buffered mesh design in 4x4, saturating at nearly the same
injection rate but maintaining a slightly lower network latency at high load.
The buffered hierarchical ring saturates at a similar point to HiRD and the
buffered mesh, but maintains a slightly lower average latency because it avoids
transfer deflections. In contrast to these high-capacity designs, the single
ring saturates at a lower injection rate, closer to the bufferless mesh design,
CHIPPER. As network size scales to 8x8, HiRD performs significantly better
relative to the meshes, because the hierarchy reduces the cross-chip latency
while preserving bisection bandwidth.

\begin{figure}[h!]
\centering
\includegraphics[width=5in]{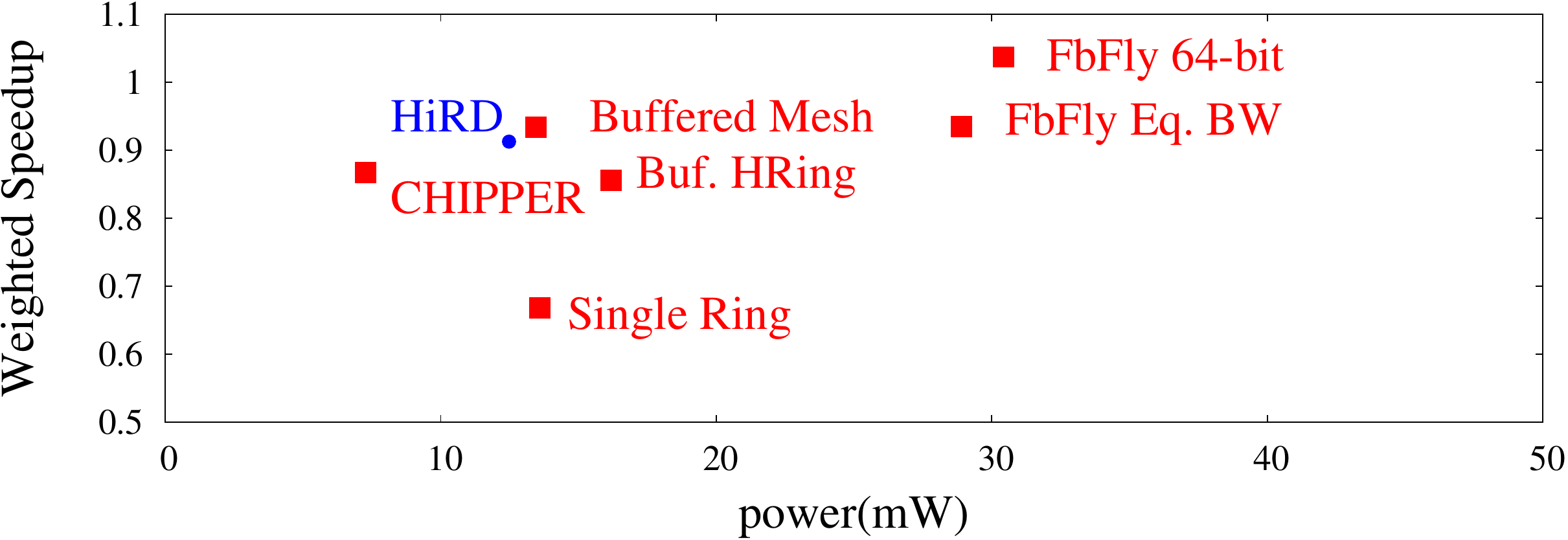}
\caption{Weighted speedup (Y) vs. power (X) for 8x8 networks.}
\label{fig:WS_power8}
\end{figure}

Figure~\ref{fig:WS_power8} shows the energy efficiency comparison for an 8x8
network. We only show HiRD with 8 bridge configuration, but varying the number
of bridge router yield similar observation as a 4x4 network. The result shows that
HiRD is more energy efficient compared to every other ring designs, and also 
provides better energy efficiency compared to most other network designs. While
CHIPPER consume roughly 33\% less power compared to HiRD. The performance of
HiRD is 5.87\% better than that of CHIPPER. In addition, comparing to the best
performing flattened butterfly network, HiRD performs roughly 10.67\% worse,
but consumes 60\% less power.

Overall, we conclude that HiRD either outperforms or is on-par with
other baseline routers with respect to performance and energy
efficiency. In cases where its efficiency is close to other designs,
the other designs require more complex routers, and HiRD's design
simplicity remains an advantage, because it reduces both
design/validation effort and router die area.

}

\section{Related Work}
\label{related}
To our knowledge, HiRD is the first hierarchical ring design that uses simple,
deflection-based ring transfers to eliminate the need for buffering within
rings while guaranteeing end-to-end packet delivery.

\noindent\textbf{Hierarchical Interconnects.} Hierarchical ring-based
interconnect was proposed in a previous line of
work~\cite{ravindran97,xiangdong95,hr-model,ravindran98,numachine,kim-hpca14}.
We have already extensively compared to past hierarchical ring proposals
qualitatively and quantitatively. The major difference between our proposal and
this previous work is that we propose deflection-based bridge routers with
minimal buffering, and node routers with no buffering. In contrast, all of
these previous works use routers with in-ring buffering, wormhole switching and
flow control. 
Concurrent works by Kim et al. propose tNoCs, hybrid packet-flit credit-based
flow control~\cite{kim-hpca14} and Clumsy Flow Control (CFC)~\cite{cfc-cal}.
However, these two designs add additional complexity because
tNoCs~\cite{kim-hpca14} requires an additional credit network to guarantee
forward progress while CFC requires coordination between cores and memory controllers. In
contrast, flow control in HiRD is lightweight (with deflection based flow
control, the Retransmit-Once mechanism, and simpler local-to-global and
global-to-local buffers). Additionally, throttling decisions in HiRD can be
made locally in each local ring as opposed to global decisions in
CFC~\cite{cfc-cal} and tNoCs~\cite{kim-hpca14}.

Udipi et al.~proposed a hierarchical topology using global and local
buses~\cite{udipi10}. While this work recognizes the benefits of hierarchy, our
design builds upon a ring-based design instead of a bus-based design because a
ring-based design provides better scalability.
Das et al.~\cite{das09} examined several hierarchical
designs, including a concentrated mesh (one mesh router shared by
several nearby nodes).

A previous system, SCI (Scalable Coherent Interface)~\cite{sci},
also uses rings, and can be configured in many topologies (including
hierarchical rings). However, to handle buffer-full conditions, SCI
NACKs and subsequently retransmits packets, whereas HiRD deflects only
single flits (within a ring), and does not require the sender to
retransmit its flits. SCI was designed for off-chip interconnect,
where tradeoffs in power and performance are very different than in
on-chip interconnects. The KSR (Kendall Square Research) machine~\cite{ksr}
uses a hierarchical ring design that resembles HiRD, yet these techniques
are not disclosed in detail and, to our knowledge, have not been 
publicly evaluated in terms of energy efficiency.

\noindent\textbf{Other Ring-based Topologies.} Spidergon~\cite{spidergon}
proposes a bidirectional ring augmented with links that directly connect nodes
opposite each other on the ring. These additional links reduce the average hop
distance for traffic. However, the cross-ring links become very long as the
ring grows, preventing scaling past a certain point, whereas our design has no
such scaling bottleneck. Octagon~\cite{octagon} forms a network by joining
Spidergon units of 8 nodes each. Units are joined by sharing a ``bridge node''
in common. Such a design scales linearly. However, it it does not make use of
hierarchy, while our design makes use of global rings to join local rings.

\noindent\textbf{Other Low Cost Router Designs.} Kim~\cite{kim09} proposes a
low-cost router design.
\ignore{that is superficially similar to our proposed node router design
routers convey traffic along rows and columns in a \emph{mesh}
without making use of crossbars, only pipeline registers and MUXes. 
Once traffic enters a row or column, it continues until it reaches its destination,
as in a ring. Traffic also transfers from a row to a column analogously to a
ring transfer in our design, using a ``turn buffer.'' 
However, because a turn
is possible at any node in a mesh, every router requires such a buffer; in
contrast, we require similar transfer buffers only at bridge routers, and their
cost is paid for by all nodes. }
However, this design is explicitly designed for meshes, hence would not be
directly usable in our ring-based design because of potential livelock as we
discussed in~\cite{hird}.
Additionally, this design does not use deflections when there is contention.
Mullins et al.~\cite{mullins04} propose a buffered mesh router with
single-cycle arbitration. 
Our work differs in that our focus is on hierarchical rings rather
than meshes. Abad et al.~\cite{rotary-router} propose the Rotary Router.
\ignore{, that consists of two independent rings that join the router's
ports and perform packet arbitration similar to standalone ring-based networks.
Both the Rotary Router and HiRD allow a packet to circle a ring again in a
``deflection'' if an ejection (ring transfer or router exit) is unsuccessful.}
Their design fundamentally differs from ours because each router has
an internal ring, and the network as a whole is a mesh. In contrast, HiRD's routers
are simpler as they are designed for hierarchical rings. 
Kodi et al.~\cite{Kodi08} propose an orthogonal
mechanism that reduces buffering by using \ignore{dual-function}links as buffer space
when necessary. Multidrop Express Channels~\cite{grot09} also provides a low cost
mechanism to connect multiple nodes using a multidrop bus without expensive router changes.

\noindent\textbf{Bufferless Mesh-based Interconnects.} While we focus
on ring-based interconnects to achieve simpler router design and lower
power, other work modifies conventional buffered mesh routers by
removing buffers and using
deflection~\cite{hotpotato,gomez08,scarab,chaosrouter,casebufferless,chipper,hat-sbac-pad,minbd,hotnets2010,sigcomm12}. 
Applying bufferless routing principles to rings leads to inherently
simpler designs, as there is only one option for deflection in a
ring (i.e., continue circulating around the ring). Other works propose
dropping packets under contention~\cite{gomez08,bless_switching}. SCARAB~\cite{scarab} adds a dedicated circuit-switch network to
send retransmit requests. Several machines such as HEP~\cite{hep}, Tera~\cite{tera}
and the Connection Machine~\cite{cm} also use deflection routing to connect
different chips.

\noindent\textbf{QoS-aware Interconnect.} Several recent works~\cite{pvc,grot2010,kilonocs,stc,aergia,maze-routing,a2c,het-nocs-dac13} introduce mechanisms
to improve QoS, fairness and delivery guarantees provided to different flows and
applications on the on-chip network. These are orthogonal to our proposal. Future works
can focus on fairness and QoS issues in the network designs we have proposed and examined.
Several techniques that provide QoS and predictability
~\cite{fairness-mem,parbs,sms,asm-micro15,mise-hpca13,dash-taco16,bliss,bliss-tpds,tcm,kim-rtas14,hui-vee,pvc,grot2010,kilonocs,stc,aergia,maze-routing,a2c,het-nocs-dac13,fst,eiman-isca11,eiman-micro09,nesbit-micro06,iyer-ics04,iyer-sigmetric07}
can be integrated to provide better QoS and fairness to a hierarchical network-on-chip.

\section{Conclusion}

We introduced \emph{HiRD}, 
a simple hierarchical ring-based NoC design. Past work has shown that a
hierarchical ring design yields good performance and scalability relative to
both a single ring and a mesh. HiRD has two new contributions: (1) a simple
router design that enables ring transfers \emph{without in-ring buffering or
flow control}, instead using limited \emph{deflections} (retries) when a flit
cannot transfer to a new ring, and (2) two \emph{guarantee mechanisms} that
ensure deterministically-guaranteed forward progress despite deflections. Our evaluations show
that HiRD enables a simpler and lower-cost implementation of a hierarchical
ring network. Although an exhaustive topology comparison is not the goal of
this work, our evaluations also show that 
HiRD is more energy-efficient than several other topologies while providing
competitive performance.
We conclude that HiRD represents a compelling interconnect design point to bring additional
scalability to existing ring-based designs at high energy efficiency.

\section*{Acknowledgments}

We thank the reviewers and SAFARI members for their feedback. We acknowledge
the support of AMD, IBM, Intel, and Qualcomm. This research was partially
supported by Intel Science and Technology Center on Cloud Computing, NSF (CCF
0953246 and CCF 1212962), and SRC. Rachata Ausavarungnirun is supported in part
by the Royal Thai Government scholarship. This article is a significantly
extended and revised version of our previous work that appeared at SBAC-PAD
2014~\cite{hird}. Our technical report~\cite{hird-safari-tr} includes results
from both this article and our previous SBAC-PAD paper~\cite{hird}.

\section*{References}

{
\bibliographystyle{abbrv}
\bibliography{paper}
}

\end{document}